\documentclass[%
 reprint,
superscriptaddress,
 amsmath,amssymb,
 aps,
pra,
floatfix,
showkeys
]{revtex4-2}

\usepackage{graphicx}
\usepackage{xcolor}
\usepackage{subcaption}
\DeclareMathOperator*{\argmin}{argmin}
\DeclareMathOperator*{\argmax}{argmax}

\begin{document}
\title{Efficient estimation of error bounds for quantum multiparametric imaging with constraints}
\author{Alexander Mikhalychev}
\affiliation{B.I. Stepanov Institute of Physics, NAS of Belarus, Nezavisimosti ave. 68-2, Minsk 220072, Belarus}
\email{mikhalychev@gmail.com}
\author{Saif Almazrouei}
\affiliation{Directed Energy Research Centre, Technology Innovation Institute, Abu Dhabi, United Arab Emirates}
\author{Svetlana Mikhalycheva}
\affiliation{B.I. Stepanov Institute of Physics, NAS of Belarus, Nezavisimosti ave. 68-2, Minsk 220072, Belarus}
\author{Abdellatif Bouchalkha}
\affiliation{Directed Energy Research Centre, Technology Innovation Institute, Abu Dhabi, United Arab Emirates}
\author{Dmitri Mogilevtsev}
\affiliation{B.I. Stepanov Institute of Physics, NAS of Belarus, Nezavisimosti ave. 68-2, Minsk, Belarus}
\author{Bobomurat Ahmedov}
\affiliation{Institute of Theoretical Physics, National University of Uzbekistan, Tashkent 100174, Uzbekistan}
\affiliation{Institute for Advanced Studies, New Uzbekistan University, Movarounnahr str. 1, Tashkent 100000, Uzbekistan}

\begin{abstract}
    Advanced super-resolution imaging techniques require specific approaches for accurate and consistent estimation of the achievable spatial resolution. Fisher information supplied to Cram{\'e}r-Rao bound (CRB) has proved to be a powerful and efficient tool for resolution analysis and optical setups optimization. However, the standard CRB is not applicable to constrained problems violating the unbiasedness condition, while such models are frequently encountered in quantum imaging of complex objects. Complementary to the existing approaches based on modifying CRB, we propose a practical algorithm for approximate construction of a modified Fisher information matrix, which takes the constraints into account and enables accurate estimation of errors for constrained problems by the standard CRB. We demonstrate the efficiency of the proposed technique by applying it to 1-, 2-, and multi-parameter model problems in quantum imaging. The approach provides quantitative explanation of previous results with successful experimental reconstruction of objects with the spatial scale smaller than the theoretical limit predicted by the standard CRB.
\end{abstract}

\maketitle

\section{Introduction}

The problem of diffraction limit of resolution in imaging was perceived and quantified more than a century ago \cite{abbe1873,rayleigh1879}. The first approaches to resolution quantification (such as Rayleigh criterion) were based on formalizing visual distinguishability of small features. While being intuitive and very fruitful in physics, the approach does not account, for example, for the influence of the shot noise on the image quality. Further, more advanced approaches based on Fourier components transfer were developed \cite{rossmann1969PSF_MTF}. Recent progress in super-resolution imaging \cite{shih_2018_introduction,magana2019quantum,schermelleh2019super,prakash2022super} forced the researchers to perceive the collected images (raw datasets) as quite abstract sources of information about the investigated sample rather than just visual replicas of it. Instead of measuring intensity of the light emitted or scattered by the sample, the super-resolution techniques are based on detection of photon coincidences \cite{shih_2018_introduction,giovannetti2009sub,xu_experimental_2015,unternahrer2018super,toninelli2019resolution}, analysis of time-dependent signal for fluctuating emitters \cite{dertinger2009fast,hess2006ultra,rust2006sub,schnitzbauer2017super,gu2013super}, or scanning with localized excitation of fluorophors \cite{hell1994breaking,hell1995ground,maurer2010far}. Since the resulting image is typically formed by elaborate processing of the raw information, the approaches to resolution quantification are to be tailored in the way suitable for taking that into account.

A powerful mathematical tool for quantification of the information content of various images (including both raw data and the processing results) is Fisher information (FI) \cite{fisher1925theory}. It describes sensitivity of the measured signal (or processed image) to variation of the sample parameters of interest. Cram{\'e}r-Rao bound (CRB) \cite{cramer1946,rao1945information} connects FI with the errors of sample parameters' estimation for optimal analysis of the acquired signal. The approach has been successfully used for resolution analysis in quantum imaging of simple \cite{ram2006beyond,motka2016optical,paur2016achieving,rehacek2017multiparameter,zhou2019quantum} and complex \cite{motka2016optical,mikhalychev2019efficiently} model objects, optimization of detection protocols \cite{tsang2016quantum,tsang2017subdiffraction,paur2018tempering,paur2019reading,len2020resolution,mikhalychev2021lost}, quantum light sources \cite{mikhalychev2019efficiently,mikhalychev2021lost}, and cumulant order in super-resolution optical fluctuation imaging (SOFI) \cite{vlasenko2020optimal,kurdzialek2021super}. Roughly speaking, resolution of a specific quantum or quantum-inspired imaging technique equals the smallest size of the sample details, such that the input signal (image) provides enough information for their reconstruction \cite{vlasenko2020optimal,mikhalychev2021lost}.

Despite of strong success in applying classical and quantum FI to theoretical investigation of quantum imaging, direct exploitation of CRB has important limitations when constraints are imposed on the considered problem. In particular, the resolution experimentally achieved in Ref. \cite{mikhalychev2019efficiently} for multiparametric quantum imaging was several times better than the theoretical limit derived from CRB without accounting for constraints. Similar effect (reduction of the estimation variance) was previously reported for experiments \cite{zhou2019quantum} and numerical simulations  \cite{tsang2016quantum,tsang2018conservative} for separation of two incoherent point-like light sources, as well as for resolution of a spectrometer \cite{motka2016optical}. 

The reason of those discrepancies is known from statistics \cite{vanTrees1968,stoica1990biased,gorman1990lower,eldar2008rethinking} and consists in the unbiasedness assumption, which is used during derivation of CRB and, generally, becomes invalid when physical constraints are taken into account. For example, for a passive non-emitting object, the absolute values of the transmission amplitude must belong to the range from 0 to 1. From physical point of view, constraints introduce additional \textit{a priori} knowledge about the sample, complementary to the information provided by the detected signal, and enable reconstruction of smaller features. Mathematically, constraints typically lead to estimation bias \cite{berger1990inadmissibility}, while biased estimators can ensure smaller mean square errors than unbiased ones \cite{stoica1990biased,eldar2004minimum,eldar2008rethinking,benHaim2009lower,motka2016optical}. 

Efficiency of CRB for resolution analysis on the one hand and typicality of constrained problems in practical imaging on the other hand motivate one to search for a way to restore applicability of CRB to such problems. One of the available approaches is based on constructing the modified CRB \cite{vanTrees1968,gorman1990lower,eldar2004minimum,eldar2008rethinking}, which explicitly takes the bias into account by incorporating its derivatives over the parameters of interest. Such approach has been successfully used for simple 1-parameter models in quantum imaging and metrology: estimation of the separation between 2 incoherent point sources \cite{tsang2016quantum, tsang2018conservative,zhou2019quantum} and inference of the wavelength by a spectrometer \cite{motka2016optical}. Minimization of the predicted estimation error over the bias allows us to construct uniform (universal) version of CRB \cite{hero1996exploring,hua1999quantitative,eldar2006uniformly}. The resulting trade-off between the bias and the variance of the estimate can serve for resolution quantification in biomicroscopy \cite{meng2004modified,meng2006design,meng2008vector}. While the approach is fruitful for derivation of general theoretical bounds on estimation errors, its practical application to quantifying information content of particular quantum imaging methods and setups can be cumbersome. Analytical calculation of neither bias gradient for a selected estimator, nor the optimal bias seems feasible for a real multiparametric problem of object reconstruction in quantum imaging \cite{mikhalychev2019efficiently}, thus leading to the necessity of time-consuming numerical Monte-Carlo simulations.

In the present contribution we propose an alternative approach, which, instead of rigorous bounds, provides only an approximate assessment of the errors and resolution, but can be efficiently applied to multiparametric problems in practice. The essence of the method is to make use of the correspondence between FI and Bayesian probabilities \cite{jaynes1996,vallisneri2008}, to modify the probability distribution according to constraints as it was done previously in our paper on quantum tomography \cite{mikhalychev2015bayesian}, and to map the new probability distribution back to FI. The modification of the probability distribution can be interpreted as taking the prior information encoded in the constraints into account. The resulting updated Fisher information matrix (FIM) already effectively includes the information about the constraints and induced estimation bias and can be put directly into the initial unmodified CRB. We apply the developed approach to multiparametric estimation problems in quantum imaging and show that it is capable of explaining and quantifying the resolution enhancement due to constraints experimentally observed in Ref. \cite{mikhalychev2019efficiently}. It is worth noting that the influence of the estimation bias on the predicted errors can be much more pronounced in quantum imaging based on measurement and analysis of photon coincidences (higher-order field correlation functions) relatively to traditional intensity-based microscopy, because FI tends to become singular if the investigated object has dark (non-transmitting) areas. We pay special attention to that issue and show that the case is still covered by the proposed approach. 

The rest of the paper is organized as follows. First, we briefly introduce the main concepts from estimation and probability theory useful for further derivation and analysis of the approach. Section~\ref{sec:constrained} is devoted to constrained estimation problems and the core procedures of our paper: regularization and correction of FIM. In Section~\ref{sec:applications}, we assess the efficiency and accuracy of the proposed approach by applying it to several model problems, ranging from very simple one to practical multiparametric imaging.

\section{Fisher information, probabilities, and resolution}

\subsection{Fisher information}

General purpose of measurements, including those used in imaging, is to estimate physical characteristics of the investigated system from the collected data \cite{lehmann1998}. For that purpose, one typically needs to construct the theoretical model of the signal $\mathbf{S}(\boldsymbol{\theta})$, where the vector $\boldsymbol{\theta} = \{\theta_\mu\}$ describes the parameters to be inferred from the experiment. Actually collected signal 
\begin{equation}
    \mathbf{Y} = \mathbf{S}(\boldsymbol{\theta}) + \boldsymbol{\varepsilon},
\end{equation}
inevitably contains noise contribution $\boldsymbol{\varepsilon}$ and, therefore, represents a particular realization of random variable(s). Statistical properties of the noise can be formalized by introduction of the likelihood $L(\mathbf{Y} | \mathbf{S})$.

The ability to reconstruct the parameters of interest from the measured data and the resulting inference error are determined by the \textit{sensitivity} of the signal to the investigated parameters. The sensitivity can be quantified by the \textit{score} function \cite{fisher1925theory}:
\begin{equation}
    s_\mu(\boldsymbol{\theta}) = \frac{\partial l(\mathbf Y | \mathbf S(\boldsymbol{\theta}))}{\partial \theta_\mu},
\end{equation}
where $l(\mathbf Y | \mathbf S(\boldsymbol{\theta})) = \log L(\mathbf Y | \mathbf S(\boldsymbol{\theta}))$ is the log-likelihood.

The covariance matrix of the score represents the Fisher information matrix (FIM) \cite{fisher1925theory}:
\begin{multline}
    \label{eq:FIM_score}
    F_{\mu \nu} = \operatorname{Cov}\left(s_\mu(\boldsymbol{\theta}), s_\nu(\boldsymbol{\theta})\right)  \\ = \operatorname{E} \left(\frac{\partial l(\mathbf Y | \mathbf S(\boldsymbol{\theta}))}{\partial \theta_\mu} \frac{\partial l(\mathbf Y | \mathbf S(\boldsymbol{\theta}))}{\partial \theta_\nu} \right),
\end{multline}
where the symbol $\operatorname{E}$ denotes mathematical expectation over the distribution of the random signal $\mathbf Y$. Roughly speaking, larger values of the FIM elements correspond to more informative measurements. If the signal $\mathbf{Y}$ takes discrete values, Eq.~(\ref{eq:FIM_score}) can be re-written in the following way convenient for direct calculation:
\begin{equation}
    \label{eq:FIM_sum}
    F_{\mu \nu} = \sum_{\mathbf{Y}} \frac{1}{L(\mathbf{Y}|\mathbf{S}(\boldsymbol{\theta}))} \frac{\partial L(\mathbf Y | \mathbf S(\boldsymbol{\theta}))}{\partial \theta_\mu} \frac{\partial L(\mathbf Y | \mathbf S(\boldsymbol{\theta}))}{\partial \theta_\nu},
\end{equation}
where the summation is performed over all possible realizations of the signal $\mathbf{Y}$.

The significance of FIM is enhanced by its usage in Cram{\'e}r-Rao bound (CRB) \cite{cramer1946,rao1945information}. If the parameters inference algorithm yields \textit{unbiased} estimate $\hat {\boldsymbol{\theta}}$, i.e., its mathematical expectation equals the true value of the parameters: $\operatorname{E}(\hat {\boldsymbol{\theta}}) = \boldsymbol{\theta}$, the covariance matrix of the estimated parameters is bounded from below by the inverse of the FIM:
\begin{equation}
    \label{eq:CRB}
    \operatorname{Cov}(\hat{\boldsymbol{\theta}},\hat{\boldsymbol{\theta}}) \ge F^{-1}.
\end{equation}
The matrix inequality $A \ge B$ means that the matrix $A - B$ is positive semidefinite. 

Equation~(\ref{eq:FIM_score}) can be also rewritten as the expectation value of the log-likelihood curvature (see e.g., Ref.~\cite{lehmann1998}):
\begin{equation}
    \label{eq:FIM-Hessian}
    F_{\mu \nu} = - \operatorname{E} \left(\frac{\partial^2 l(\mathbf Y | \mathbf S(\boldsymbol{\theta}))}{\partial \theta_\mu \partial \theta_\nu} \right).
\end{equation}
In the limit of high signal-to-noise ratio (SNR) typical for strong signals, that representation implies that FIM represents Hessian of maximum-likelihood estimate (MLE, formulated as finding $\hat{\boldsymbol{\theta}}$ maximizing the log-likelihood $l(\mathbf{Y},\mathbf{S}(\hat{\boldsymbol{\theta}}))$ for the specific collected dataset $\mathbf{Y}$) and characterizes scattering of the estimation results for multiple repetitions of the experiment \cite{vallisneri2008}.

Alternatively, one can adhere to Bayesian approach \cite{koch2007} and consider posterior probability $p(\boldsymbol{\theta} | \mathbf Y)$ of the parameters vector $\boldsymbol{\theta}$, conditioned by the measured dataset $\mathbf Y$. For a uniform prior probability distribution (absence of useful information about the investigated parameters before the measurement) and high SNR, the FIM $F$ describes the shape of the posterior probability distribution near the point  $\hat{\boldsymbol{\theta}}_{ML}$ of the likelihood maximum \cite{jaynes1996,vallisneri2008}:
\begin{multline}
    \label{eq:FIM-posterior}
    p(\boldsymbol{\theta} | \mathbf Y) \propto L(\mathbf Y | \mathbf S(\boldsymbol{\theta}))  \\ \propto \exp \left[ -\frac{1}{2} \left(\boldsymbol{\theta} - \hat{\boldsymbol{\theta}}_{ML}\right) ^T F \left(\boldsymbol{\theta} -  \hat{\boldsymbol{\theta}}_{ML}\right) \right].
\end{multline}

The application of FI in terms of CRB is directly related to accuracy and resolution in imaging, while its Bayesian interpretation is helpful for our development of the constraints treatment approach. Some insights into relations between likelihood, distribution of point estimation results, and FI can also be gained from geometrical considerations presented in Appendix~\ref{app:geometry of FI}.

It is also worth mentioning \textit{quantum} Fisher information (QFI) \cite{helstrom1969quantum,holevo1982probabilistic,paris2009quantum}, which generalizes the concept of classical FI and corresponds to the maximum of classical FI over all physically valid measurements. QFI represents a powerful tool for theoretical analysis of relatively simple systems \cite{paur2016achieving,rehacek2017multiparameter,tsang2017subdiffraction,tsang2016quantum,tsang2015quantum,nair2016far,dutton2019attaining,demkowicz2020multi} and can be applied even to practical multiparametric problems \cite{lupo2020quantum,zhou2019modern,bisketzi2019quantum}, but may require cumbersome calculations in that case. Due to the same structure and similar interpretation of classical FI and QFI, our approach remains applicable to QFI as well. However, in the current work we focus on practical applications considering quantum imaging with a ﬁxed (not optimized) type of measurement and limit our analysis to classical FI.

\subsection{Fisher information in quantum imaging}

In quantum imaging, the measured signal $\mathbf S(\boldsymbol{\theta})$ is typically represented by field correlation functions \cite{shih_2018_introduction}
\begin{multline}
    \label{eq:Gn}
    G^{(n)}(\mathbf{r}_1, t_1; \ldots ; \mathbf{r}_n, t_n)  = \Bigl \langle E^{(-)}(\mathbf r_1, t_1) \cdots E^{(-)}(\mathbf r_n, t_n) \\ \times E^{(+)}(\mathbf r_n, t_n) \cdots E^{(+)}(\mathbf r_1, t_1) \Bigr\rangle,
\end{multline}
where $E^{(\pm)}(\mathbf r, t)$ are positive(negative)-frequency field operators. Namely, the signal $\mathbf{S}(\boldsymbol{\theta}) = (S_1(\boldsymbol{\theta}), S_2(\boldsymbol{\theta}), \ldots)$ is composed of the components
\begin{equation}
    S_i(\boldsymbol{\theta}) \propto G^{(n)}(\mathbf{r}_1^{(i)}, t_1^{(i)}; \ldots ; \mathbf{r}_n^{(i)}, t_n^{(i)})
\end{equation}
defined by specific arguments of the correlation function.

From practical point of view, the $n$-th order correlation function $G^{(n)}$ describes the rate of $n$-photon coincidence events for specified detection positions and time delays. The signal component $S_i(\boldsymbol{\theta})$ describes the mean number of the corresponding coincidence events, while the component $Y_i$ of a random signal realization $\mathbf{Y} = (Y_1, Y_2, \ldots)$ equals the realized (integer) number of such events.

Commonly, the detection events can be treated as independent ones and the likelihood function corresponds to independent random variables $Y_i$ obeying Poisson distribution with the mean values $S_i(\boldsymbol{\theta})$:
\begin{equation}
    \label{eq:Poisson multivariate likelihood}
    L(\mathbf{Y}|\mathbf{S}(\boldsymbol{\theta})) = \prod_i L(Y_i | S_i(\boldsymbol{\theta})),
\end{equation}
where
\begin{equation}
    \label{eq:Poisson likelihood}
    L(Y|S(\boldsymbol{\theta})) = \frac{[S(\boldsymbol{\theta})]^Y}{Y!} e^{-S(\boldsymbol{\theta})}
\end{equation}
is the likelihood function for a single-variate Poisson distribution. 

In that specific, but commonly encountered case the expression (\ref{eq:FIM_sum}) can be simplified and takes the following form (see Refs.~\cite{len2020resolution,tsang2017subdiffraction,mikhalychev2021fisher,mikhalychev2021lost} and Appendix~\ref{app:Poisson FIM}):
\begin{equation}
    \label{eq:FIM Poisson}
    F_{\mu\nu} = \sum_i \frac{1}{S_i(\boldsymbol{\theta})} \frac{\partial S_i(\boldsymbol{\theta})}{\partial \theta_\mu} \frac{\partial S_i(\boldsymbol{\theta})}{\partial \theta_\nu}.
\end{equation}
It is worth noting that the structure of Eqs.~(\ref{eq:FIM_sum}) and (\ref{eq:FIM Poisson}) for the FIM elements $F_{\mu\nu}$ is very similar, while their meaning is quite different. The general expression~(\ref{eq:FIM_sum}) is constructed from probabilities of all possible (mutually exclusive) measurement outcomes $\mathbf{Y}$, and the normalization $\sum _ {\mathbf{Y}} L(\mathbf{Y}|\mathbf{S}(\boldsymbol{\theta})) = 1$ holds. In Eq.~(\ref{eq:FIM Poisson}), the summation index $i$ enumerates different signal components measured simultaneously in one experiment. The quantities $S_i(\boldsymbol{\theta})$ represent the expectation values of the events numbers and are not subjected to any specific kind of normalization.

\begin{figure}
    \centering
    \includegraphics[width=0.65\linewidth]{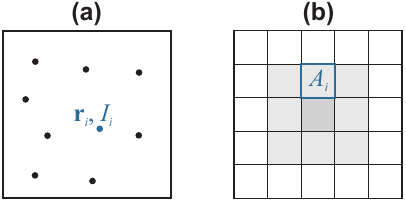}
    \caption{Problem parameterization in quantum imaging: (a) by positions $\mathbf r_i$ and brightnesses $I_i$ of point emitters; (b) by transmission amplitudes $A_i$ of pixels.}
    \label{fig:parameterization}
\end{figure}

The set of inferred parameters $\boldsymbol{\theta}$ is determined by the model of the investigated object. In quantum imaging, the model is typically defined either as a set of discrete point emitters, parameterized by their positions $\mathbf r_i$ and brightness $I_i$ (Fig.~\ref{fig:parameterization}(a)) \cite{ram2006beyond,motka2016optical,paur2016achieving,rehacek2017multiparameter,zhou2019quantum,tsang2016quantum,tsang2017subdiffraction,paur2018tempering,paur2019reading,len2020resolution,vlasenko2020optimal,zhou2019modern}, or as a pixelized mask, characterized by the transmission amplitudes of the pixels $A_i$ (Fig.~\ref{fig:parameterization}(b)) \cite{giovannetti2009sub,lemos2014quantum,pushkina2021superresolution,unternahrer2018super,xu_experimental_2015,mikhalychev2019efficiently,mikhalychev2021lost}.

FIM itself provides valuable information about the problem structure --- parameters correlations and inaccuracies (see e.g. \cite{rehacek2017multiparameter,bisketzi2019quantum,mikhalychev2019efficiently,mikhalychev2021fisher}), since variance and covariances of the parameters are related to the inverse of FIM by CRB. For various optimization purposes, as well as for quantification of optical resolution, it is also useful to construct a scalar real-valued characteristic providing aggregate description of the accuracy in a multiparametric problem. There exist several common choices of such scalar quantity \cite{kiefer1959,asprey2002,banga2002,faller2003}: the volume of the joint confidence region defined by CRB ($D$-optimality --- minimization of the determinant of the inverse of FIM: $\min \operatorname{det} F^{-1}$); the major axis of the joint confidence region ($E$-optimality --- minimization of the maximal eigenvalue of the inverse of FIM: $\min \lambda_{max}(F^{-1})$); the perimeter of the enclosing box of the joint confidence region ($A$-optimality --- minimization of the trace of the inverse of FIM: $\min \operatorname{Tr} F^{-1}$). The latter choice has already been applied to quantum imaging \cite{mikhalychev2019efficiently,vlasenko2020optimal,mikhalychev2021lost} by using the lower bound of total variance
\begin{equation}
    \label{eq:Delta_squared_def}
    \Delta^2 \equiv \operatorname{Tr}F^{-1} \le \sum_\mu \operatorname{Var}\theta_\mu 
\end{equation}
as a measure of diffraction-caused inaccuracy.

FI represents a powerful and useful tool for resolution quantification in imaging. If inference of the positions (separation) of point emitters is considered, CRB directly provides a resolution measure: the predicted (minimal) inaccuracy of the position reconstruction quantifies the resolving power of the microscope \cite{ram2006beyond,motka2016optical,paur2016achieving,zhou2019quantum,tsang2016quantum,paur2018tempering,paur2019reading}. For analysis of pixelized images, one can start from the notion that a feature is resolved if the acquired image (dataset) contains enough information for inference of the feature parameters with certain requested accuracy \cite{ram2006beyond,motka2016optical,vlasenko2020optimal,mikhalychev2021lost}. In particular, one can define the spatial resolution as the minimal spatial scale of the problem (or feature size), for which the quantity defined by Eq.~(\ref{eq:Delta_squared_def}) is lower than certain threshold value determined by further exploitation purposes of the imaging results \cite{vlasenko2020optimal,mikhalychev2021lost}: $\Delta^2 \le \Delta^2_\text{threshold}$.

The close connection of FI to achievable resolution allows us to use it for optimization of the imaging setup. Increase of FI, including removal of ``Rayleigh's curse'' in simple cases, indicated apparent advantages of detection protocols based on spatial modes demultiplexing \cite{tsang2016quantum,tsang2017subdiffraction,len2020resolution,karuseichyk2022resolving,rouviere2024ultra}, modification of point-spread function (PSF) \cite{paur2018tempering,paur2019reading}, or probabilistic transformation of an entangled light state \cite{mikhalychev2021lost}. Analysis of FI proved to be useful in optimization of quantum light sources \cite{mikhalychev2019efficiently,mikhalychev2021lost} and selection of the optimal cumulant order in super-resolution optical fluctuation imaging (SOFI) \cite{vlasenko2020optimal,kurdzialek2021super}.

In both object parameterizations discussed above, some basic constraints can limit the range of meaningful values for the parameters. For example, brightness of any emitter or the absolute value of the separation between two emitters should be non-negative ($I_i \ge 0$, $\Delta r \equiv |\mathbf r_1 - \mathbf r_2| \ge 0$). For a passive transmitting object, the absolute value of a pixel transmission amplitude cannot exceed 1: $|A_i| \le 1$. In a commonly encountered case of an object with a real-valued transmission amplitude (see e.g. Refs.~\cite{agafonov2009high,chen_arbitrary-order_2010,toninelli2019resolution,unternahrer2018super,giovannetti2009sub,moreau_Resolution_2018,xu_experimental_2015}), the constraint takes the form 
\begin{equation}
    \label{eq:constraint_on_amplitude}
    0 \le A_i \le 1.
\end{equation}

A remarkable illustration of the constraints influence on the practically achievable spatial resolution was provided by the experiment reported in Ref.~\cite{mikhalychev2019efficiently}. A number of pixelized objects (1-dimensional sets of slits and 2-dimensional masks) were illuminated by biphotons and pseudothermal light and the second- and third-order correlation functions were recorded. Quite unexpectedly, actually achieved resolution was several times higher than the theoretical limit, predicted without taking the constraints into account. Namely, the resolution limit (minimal pixel size $d_\text{min}$ defined from FI according to the predicted inference accuracy for the pixels' transmission amplitudes \cite{mikhalychev2021lost}) was approximately a half of the Rayleigh limit $d_\text{R}$ for the used optical system: $d_\text{min} \approx (0.4\div 0.5) d_\text{R}$. The experimental datasets allowed successful object features reconstruction for the parameterization with a  noticeably smaller pixel size: $d \approx (0.25 \div 0.28) d_\text{R} < d_\text{min}$. 

Those counter-intuitive results were explained by a specific feature (still, quite common for model experiments) of the imaged objects: the actual transmission amplitude of their pixels took binary values $A_i = 0$ or 1. Therefore, the constraints (\ref{eq:constraint_on_amplitude}) were active (i.e., influenced the reconstruction process) and provided additional information about the sample. It is worth noting, that the information about such binary type of the objects was not implied \textit{a priori} and was not used to assist the reconstruction algorithm. The connection between the resolution enhancement and the constraints was verified by Monte-Carlo simulations and numerical estimation of the inference accuracy. The achievable resolution (minimal pixel size for reliable reconstruction of the transmission amplitudes) was approximately twice better for black-and-white model objects (with active constraints) than for the corresponding gray objects (when the constraints are inactive because the values are far from 0 and 1). However, a way to quantitative theoretical FI-based prediction of such resolution improvement was not considered in Ref.~\cite{mikhalychev2019efficiently}. In the current contribution, we aim to fill that gap.

\section{Constrained estimation and Fisher information}
\label{sec:constrained}
\subsection{Influence of constraints on parameter estimation}

When the estimation formalism is applied to real-world tasks, the true values of the parameter may be constrained by a specific physical domain: $\boldsymbol{\theta} \in \Omega$. In that case, it is reasonable to impose the constraints on the estimated values as well: $\hat{\boldsymbol{\theta}} \in \Omega$. If the constraints are active (i.e., the true values of the parameters are within the estimation inaccuracy from the boundaries of the region $\Omega$), the relation between FI and the estimation errors becomes more complicated than prescribed by CRB (Eq.~(\ref{eq:CRB})). The constrained estimation typically violates unbiasedness condition \cite{berger1990inadmissibility}, required for validity of CRB. 

In Bayesian treatment of the problem, the constraints induce nontrivial prior probability: $p_0(\boldsymbol{\theta}) = 1 / | \Omega |$ for $\boldsymbol{\theta} \in \Omega$ and 0 otherwise. Since prior probabilities are not taken into account by FI, it does not describe the posterior probability by Eq.~(\ref{eq:FIM-posterior}) correctly for constraint problems. 

\begin{figure}
    \centering
    \includegraphics[width=1.0\linewidth]{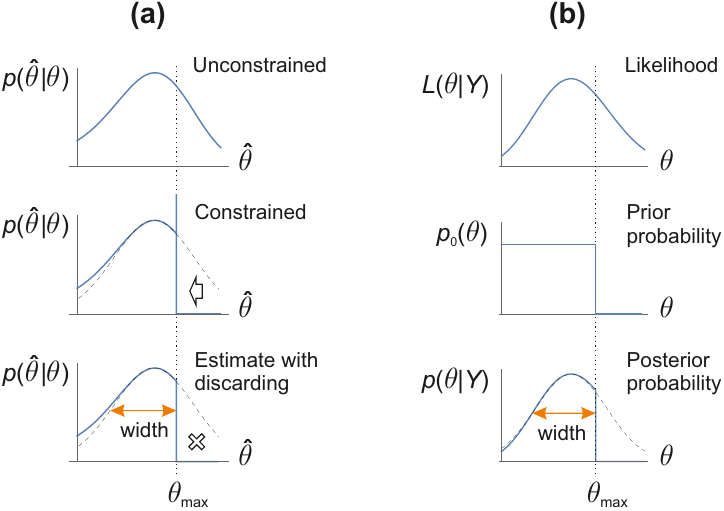}
    \caption{Influence of a constraint $\theta \le \theta_\text{max}$ on the estimate distribution: (a) distribution of the estimated parameter value $\hat \theta$ for its true value $\theta$ (from top to bottom --- unconstrained estimate, constrained estimate, and formally introduced estimate with discarded nonphysical values); (b) Bayesian posterior probability for the parameter $\theta$ for a given signal value $Y$ (from top to bottom --- unconstrained likelihood, prior probability distribution describing the constraint, and the resulting posterior probability). Dashed lines show Gaussian fit of the unconstrained distributions. The resulting width after the constraint application is smaller than the one predicted by FI.}
    \label{fig:constrained 1D}
\end{figure}

The influence of the constraints on the distribution of estimated parameter values in 1-parameter case is shown in Fig.~\ref{fig:constrained 1D} (provided just as an illustration without connection to any specific physical model). It is worth noting, that Bayesian ``cutting'' of the posterior probability distribution by setting zero prior probability for nonphysical values can be formally reproduced in the distribution of point estimation results by performing unconstrained estimation and subsequent discarding of nonphysical values $\hat{\boldsymbol{\theta}}_\text{unconstr.} \not\in \Omega$. In high-SNR case, the resulting distributions will have practically the same width. While that estimate with the values discarding does not follow typical practical procedures of parameter inference, it is useful for getting some insights for the development of the FIM modification approach.

\begin{figure*}
    \centering
    \includegraphics[width=0.9\linewidth]{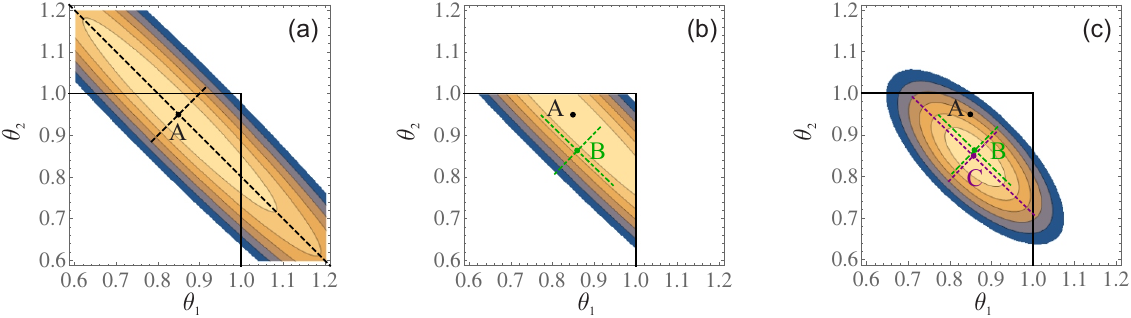}
    \caption{Influence of constraints $\theta_1 \le 1$ and $\theta_2 \le 1$ on the distribution of parameters' estimates for a model 2-parameter case: probability distribution for an unconstrained problem (a), constrained distribution corresponding to posterior probability for Bayesian treatment or to the formal distribution of the estimated parameters values after discarding nonphysical results (b), and Gaussian approximation of the constrained distribution according to Subsection~\ref{subsec:effective FI} (c). Capital letters A, B, and C indicate the mean values for the probability distributions shown in panels (a), (b), and (c) respectively. Dashed lines correspond to the eigenvectors of the covariance matrix and indicate the corresponding widths of the distributions. The constrained estimate is biased (the point B is shifted from the point A indicating the true values of the parameters $\theta_1$ and $\theta_2$). Errors of the parameters' estimation become approximately 6 times smaller when the physical constraints are taken into account.}
    \label{fig:constrained 2D}
\end{figure*}

In 1-parameter case, the constraint-induced decrease of the parameter reconstruction error (the width of the estimate distribution) is quite limited. Taking into account that the true value $\theta \in \Omega$ belongs to the physical domain and assuming that the distribution $p(\hat{\theta} | \theta)$ is symmetric around the true value $\theta$, one can easily see that the width is minimal for $\theta$ having a boundary value ($\theta = \theta _ \text{max}$ for the case shown in Fig.~\ref{fig:constrained 1D}) and reaches 1/2 of the initial width of the unconstrained distribution. The effect can be much more pronounced for a multi-parameter problem \cite{mikhalychev2019efficiently}. Fig.~\ref{fig:constrained 2D} illustrates the statement for a model 2-parameter problem, where the error is reduced by the factor of approximately 6.

\subsection{Effective Fisher information matrix for constrained estimation}
\label{subsec:effective FI}

For unconstrained problems, FIM can be used for efficient prediction of experimental inference inaccuracies via CRB (Eq.~(\ref{eq:CRB})). In high-SNR regime, FIM also represents the quadratic-form kernel for Gaussian approximation of the estimated parameters distribution (Eq.~(\ref{eq:FIM-posterior})). Mathematically, the two meanings of FIM are ensured by the relation between the quadratic form and the covariance matrix of a multivariate normal distribution:
\begin{equation}
    \label{eq:quadratic form covariance}
    p(\hat{\boldsymbol{\theta}}) \propto \exp\left[ - \frac{1}{2} \Delta \hat{\boldsymbol{\theta}}^T F \Delta \hat{\boldsymbol{\theta}} \right] \Rightarrow \operatorname{Cov}(\hat{\boldsymbol{\theta}}, \hat{\boldsymbol{\theta}}) = F^{-1},
\end{equation}
where $\Delta \hat{\boldsymbol{\theta}} = \hat{\boldsymbol{\theta}} - \operatorname{E}(\hat{\boldsymbol{\theta}})$.

When the constraints are applied to the problem, the basic CRB becomes invalid (due to bias of the estimate) while usage of its modified versions \cite{vanTrees1968,gorman1990lower,eldar2004minimum,eldar2008rethinking} is typically computationally expensive. On the other hand, transformation of probability distributions $p(\hat{\boldsymbol{\theta}}) \mapsto p_\text{constr.}(\hat{\boldsymbol{\theta}})$ caused by application of the constraints is quite straightforward, as discussed above. 

\begin{figure}
    \centering
    \includegraphics[width=1.0\linewidth]{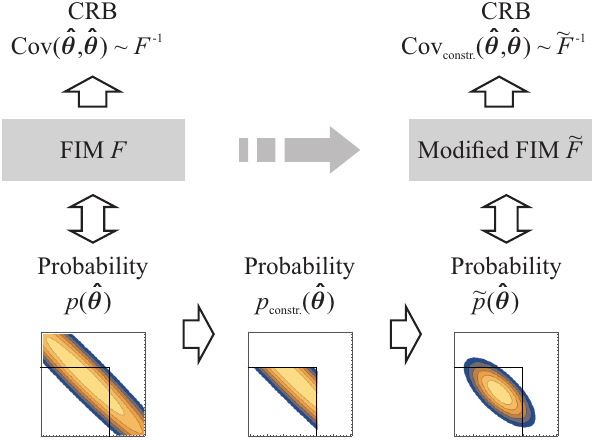}
    \caption{Scheme of taking constraints into account by construction of modified FIM. The basic FIM $F$ corresponds to the unconstrained problem and, in high-SNR regime, represents the quadratic-form kernel for the probability distribution $p(\hat{\boldsymbol{\theta}})$ of unconstrained parameter estimates. Then, the constraints are applied to the probability distribution and the result is approximated by a multivariate normal distribution $\tilde p(\hat{\boldsymbol{\theta}})$. The kernel $\tilde F$ of that distribution's quadratic form can be interpreted as the modified FIM since it is related to the covariance of the constrained parameter estimates through usual CRB.}
    \label{fig:modified FIM scheme}
\end{figure}

From practical point of view, one is interested in errors of the parameters estimation, namely in the covariance matrix $\operatorname{Cov}_\text{constr.}(\hat{\boldsymbol{\theta}}, \hat{\boldsymbol{\theta}})$ for the constrained distribution. If the distribution $p_\text{constr.}(\hat{\boldsymbol{\theta}})$ can be approximated by a multivariate normal distribution $\tilde p (\hat{\boldsymbol{\theta}})$, the relation similar to Eq.~(\ref{eq:quadratic form covariance}) will hold:
\begin{equation}
    \label{eq:quadratic form covariance approx}
    \tilde p(\hat{\boldsymbol{\theta}}) \propto \exp\left[ - \frac{1}{2} \Delta \hat{\boldsymbol{\theta}}^T \tilde F \Delta \hat{\boldsymbol{\theta}} \right] \Rightarrow \operatorname{Cov}_\text{constr.}(\hat{\boldsymbol{\theta}}, \hat{\boldsymbol{\theta}}) \approx \tilde F^{-1},
\end{equation}
where $\Delta \hat{\boldsymbol{\theta}} = \hat{\boldsymbol{\theta}} - \operatorname{E}_\text{constr.}(\hat{\boldsymbol{\theta}})$. The quadratic form $\tilde F$ of the probability distribution $\tilde p(\hat{\boldsymbol{\theta}})$ can be interpreted as the FIM for the constrained problem and quantifies the error of constrained parameters estimation via usual (unmodified) CRB defined by Eq.~(\ref{eq:CRB}). The transition from the initial FIM $F$ to the modified FIM $\tilde F$ is shown schematically in Fig. \ref{fig:modified FIM scheme}. Since the proposed procedure of FIM modification preserves the form of CRB for constrained problems, further analysis methods based on CRB (e.g., quantification of resolution \cite{mikhalychev2021lost} or optimization of the measurement setup \cite{mikhalychev2019efficiently}) remain valid.

Computationally efficient procedure of transforming $p(\hat{\boldsymbol{\theta}})$ into $\tilde p(\hat{\boldsymbol{\theta}})$ has been proposed in Ref.~\cite{mikhalychev2015bayesian} in the connection with adaptive quantum tomography with Bayesian update of the knowledge about the analyzed quantum state. The constraints are assumed to be linear and have the form $\mathbf{a}_j ^T \boldsymbol{\theta} \le b_j$. The idea of the approach consists in iterative shrinking of the probability distribution according to the current most severely violated constraint. 

Let the probability distribution obtained after the $i$-th iteration be described as
\begin{equation}
    p^{(i)}(\hat{\boldsymbol{\theta}}) \propto \exp\left[ - \frac{1}{2} \Delta \boldsymbol{\theta}^T F^{(i)} \Delta \boldsymbol{\theta} \right], \quad \Delta \boldsymbol{\theta} = \hat{\boldsymbol{\theta}} -\boldsymbol{\theta}^{(i)}.
\end{equation}
Then, the $(i+1)$-th iteration transforms $p^{(i)}(\hat{\boldsymbol{\theta}})$ into $p^{(i + 1)}(\hat{\boldsymbol{\theta}})$ and includes the following steps \cite{mikhalychev2015bayesian} (see Appendix~\ref{app:Iterative algorithm} for a visual illustration and additional expressions):
\begin{itemize}
    \item Construct a matrix  $T = T^T = (F^{(i)})^{1/2}$ and apply a linear coordinate transformation $\hat{\boldsymbol{\theta}} \mapsto \boldsymbol{\theta}' = T (\hat{\boldsymbol{\theta}} - \boldsymbol{\theta}^{(i)})$, such that the transformed distribution has zero mean ${\boldsymbol{\theta}^{(i)}}' = 0$ and the quadratic-form kernel represented by the unit matrix: ${F^{(i)}}' = T^{-1} F^{(i)} T^{-1}  = 1$. The transformation is non-degenerate, if $F^{(i)}$ is not singular.
    \item Recalculate all the constraints to the new coordinates as $\mathbf{a}_j \mapsto \mathbf{a}_j' = \mathbf{a}_j T^{-1}$, $b_j \mapsto b_j' = b_j - \mathbf{a}_j^T \boldsymbol{\theta}^{(i)}$.
    \item Find the index of the most severely violated constraint: $j = \argmin_j (b_j' / |\mathbf{a}_j'|)$.
    \item Shrink the distribution in the direction $\mathbf{d} = \mathbf{a}_j' / |\mathbf{a}_j'|$ corresponding to the found constraint: ${F^{(i)}}' \mapsto {F^{(i)}}'' + \xi \mathbf{d}\mathbf{d}^T$, ${\boldsymbol{\theta}^{(i)}}' \mapsto {\boldsymbol{\theta}^{(i)}}'' = -\delta \mathbf{d}$. Explicit expressions for $\delta$ and $\xi$ are listed in Appendix~\ref{app:Iterative algorithm}.
    \item Apply the inverse transform to get the resulting probability distribution after the current iteration: $F^{(i+1)} = T {F^{(i)}}'' T = F^{(i)} + \xi T \mathbf{d} \mathbf{d}^T T$, $\boldsymbol{\theta}^{(i+1)} = \boldsymbol{\theta}^{(i)} - \delta T^{-1} \mathbf{d}$.
\end{itemize}

\subsection{Ill-defined Fisher information matrix for dark objects}
\label{subsec:ill-defined}

Both, application of CRB (\ref{eq:CRB}) to the unmodified FIM and implementation of the proposed iterative correction require FIM to be non-singular in order to get meaningful results. However, multi-parameter estimation problem in quantum imaging can yield an ill-defined FIM if the object contains a dark (non-transparent) region. 

In particular, for a uniform object with the transmission amplitude $A$, the $n$-th order correlation function $G^{(n)}$, defined by Eq.~(\ref{eq:Gn}), is polynomial relatively to $A$: $G^{(n)} \propto A^{2n}$. Then, FI, calculated for the problem parameter $A$, scales as $F \propto A^{2(n-1)}$. For conventional intensity-based imaging ($n=1$), the singularity is absent: $2(n-1) = 0$ and $F$ remains non-zero for $A = 0$. However, for detection of correlation functions with $n \ge 2$, one has $F = 0$ for $A = 0$ (quite similarly to ``Rayleigh's curse'' for inference of two point light sources separation \cite{tsang2016quantum}).

As another illustrative example of quantum imaging, one can consider illumination of the object by ideally correlated biphotons. The second-order correlation function can be represented as \cite{shih_2018_introduction}
\begin{multline}
    G^{(2)}(\mathbf{r}_1, \mathbf{r}_2) \propto |\Psi(\mathbf{r}_1, \mathbf{r}_2)|^2 \\ \equiv \left| \int d^2 \mathbf{s} A^2(\mathbf{s}) h(\mathbf{s}, \mathbf{r}_1) h(\mathbf{s}, \mathbf{r}_2)\right|^2
\end{multline}
where $A(\mathbf{s})$ is the transmission amplitude of the object at the point with the transverse position $\mathbf{s}$, and $h(\mathbf{s}, \mathbf{r})$ is the PSF (Green's function) of the optical system. 

For a pixelized object, an array detector, and a general case of non-ideally correlated biphotons, the second order correlation between the $i$-th and $j$-th pixels of the detector can be expressed as \cite{mikhalychev2019efficiently}
\begin{equation}
    \label{eq:G2 discretized}
    G^{(2)}_{ij} \propto |\Psi_{ij}|^2 = \biggl| \sum_{m,l} D^{(ij)}_{ml} A_m A_l \biggr|^2
\end{equation}
with certain coefficients $D^{(ij)}_{ml}$ (real-valued for simple models \cite{mikhalychev2019efficiently}). If biphotons are ideally correlated, the additional relation $D^{(ij)}_{ml} = 0$ holds for $m \ne l$.  According to Eq.~(\ref{eq:FIM Poisson}), diagonal elements of FIM are proportional to 
\begin{equation}
    F_{mm} \propto A_m^2 \sum_{i,j} \left( D^{(ij)}_{mm} \right)^2 \propto A_m^2.
\end{equation}
for the considered parameterization. Therefore, the adverse scaling of FIM for $A_m \rightarrow 0$ remains the same as for the toy example of a uniform object.

\subsection{Regularization of ill-defined Fisher information matrix}
\label{subsec:regularization}

An insight to the reasons of FIM singularity and inapplicability of CRB for quantum imaging of dark objects can be gained from Eqs.~(\ref{eq:FIM-Hessian}) and (\ref{eq:FIM-posterior}). Roughly speaking, the local properties of the log-likelihood, characterized by its second derivative over the parameters, are used to estimate the distribution shape and to quantify its width. However, such local properties can be non-representative if the distribution shape is far from the assumed one.

To assist our intuition, it is worth considering a simple problem resembling the discussed behavior of FI. Let us suppose we are given a profile $y(x)$ with the maximum at $x = 0$ and need to estimate the width of that peak. If the peak has Gaussian shape ($y(x) \propto e^{-x^2 / (2\sigma^2)}$), the half-width $\sigma$ can be estimated by calculating the second derivative of the profile logarithm:
\begin{equation}
    \label{eq:width by D2}
    \sigma = \frac{1}{\sqrt{|Y^{(2)}(0)|}},\quad Y^{(2)}(x) \equiv \frac {d^2}{dx^2} \left( \log y \right).
\end{equation}

\begin{figure}
    \centering
    \includegraphics[width=1.0\linewidth]{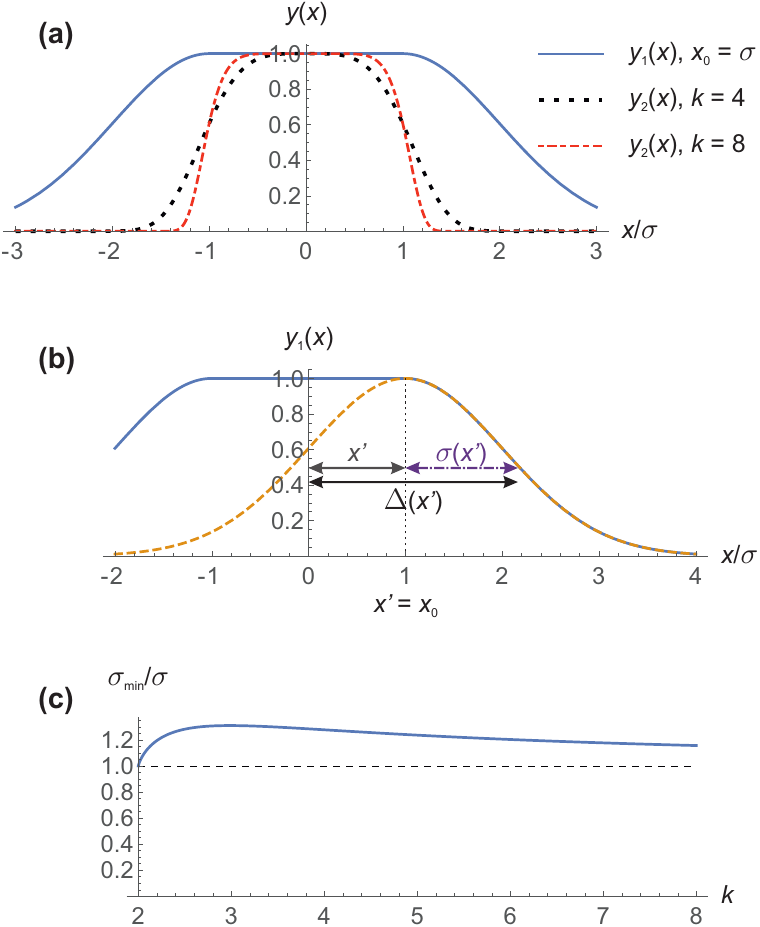}
    \caption{Width estimation by the second derivative. (a)~Model profiles defined in the text (blue solid line --- $y_1(x)$ with $x_0 = \sigma$; black dotted line --- $y_2(x)$ with $k = 4$; red dashed line --- $y_2(x)$ with $k = 8$). (b)~Width estimation by ``scanning'' the profile by a Gaussian-shape peak: total half-width $\Delta(x')$ includes the half-width $\sigma(x')$ estimated at the shifted position $x'$ and the displacement $x'$. (c)~Half-width estimate $\sigma_\text{min}$ for $y_2(x)$ according to the proposed approach normalized by the true half-width $\sigma$ of the profile (at the height $e^{-1/2}$).}
    \label{fig:width regularization}
\end{figure}

Now, let us consider the profiles $y_1 = e^{-\max(0, (|x|-x_0))^2 / (2\sigma^2)}$ and $y_2 = e^{-|x|^k  / (2\sigma^k)}$, $k > 2$ (Fig.~\ref{fig:width regularization}(a)). In both cases, we have $Y^{(2)}(0) = 0$ and Eq.~(\ref{eq:width by D2}) is inapplicable for estimation of $\sigma$, while the actual width remains finite. To get meaningful width estimates, one can probe the slope of the peak by calculating $Y^{(2)}(x')$ for shifted positions $x' \ne 0$ and construct the quantities
\begin{equation}
    \label{eq:shifted width}
    \Delta(x') = |x'| + \sigma(x') \equiv |x'| + \bigl|Y^{(2)}x')\bigr|^{-1/2}.
\end{equation}
Geometrically, we fit different parts of the peak slope by a Gaussian-shape peak and estimate the resulting width (Fig.~\ref{fig:width regularization}(b)). For the profile $y_1(x)$, the value
\begin{equation}
    \label{eq:width estimate by min}
    \sigma_\text{min} = \min_{x'} \Delta(x')
\end{equation}
will be exactly the half-width of the peak (Fig.~\ref{fig:width regularization}(b)). For the function $y_2(x)$, Eq.~(\ref{eq:width estimate by min}) provides an approximate result, which, however, is still quite accurate (see Fig.~\ref{fig:width regularization}(c) and Appendix~\ref{app:width estimation}).

One should treat the discussed illustrative example with caution: the toy example deals with a continuous dependence $y(x)$, while the actually detected signal $\mathbf{Y}$ in quantum imaging typically represents a vector of discrete integer-valued random variables. Moreover, it is the discretness 
of the detected signal that leads to singularity of FIM for a dark object (Appendix~\ref{app:discretness}). Still, the gained insights can be used for regularization of FIM.

To design the regularization procedure for FIM, we start from considering 1-parameter case. When the error $\Delta$ of estimating the parameter $\theta$ is analyzed on the base of CRB, FI $F(\theta)$ (which, in the case of a single parameter, is a scalar --- $1 \times 1$ matrix) plays the role similar to the quantity $Y^{(2)}(x)$, introduced in Eq.~(\ref{eq:width by D2}):
\begin{equation}
    \label{eq:width by FIM 1D}
    \Delta \sim \frac{1}{\sqrt{F(\theta)}}.
\end{equation}
That analogy is also supported by definition of FIM according to Eq.~(\ref{eq:FIM-Hessian}). 

In the case of a dark object ($F \approx 0$), Eq.~(\ref{eq:width by FIM 1D}) strongly overestimates the error, since the second derivative is not representative enough for the probability distribution shape. Following the idea of Eqs.~(\ref{eq:shifted width}) and (\ref{eq:width estimate by min}), one can define
\begin{equation}
    \label{eq:Delta theta shifted}
    \Delta(\theta') = |\theta' - \theta| + F^{-1/2}(\theta')
\end{equation}
and estimate the inaccuracy as 
\begin{equation}
    \label{eq:Delta min}
    \Delta_\text{min} = \min_{\theta'} \Delta(\theta').
\end{equation}
The approach is illustrated by  Fig.~\ref{fig:regularization geometry} in Appendix~\ref{app:geometry of FI}.

Now, we aim at constructing a modified FI $\tilde F (\theta)$ such that the initial form of the relation $\Delta = \Delta_\text{min} \sim  1/\sqrt{\tilde F(\theta)}$ holds. This requirement can be fulfilled by defining
\begin{equation}
    \label{eq:regularized FIM 1D}
    \tilde F(\theta) = \frac{1}{\Delta_\text{min}^2} =  \max_{\theta'} \frac{F(\theta')}{\left(1 + |\theta' - \theta| \sqrt{F(\theta')}\right)^2}.
\end{equation}

In a general multi-parameter case, one can apply the discussed regularization along main axes of FIM independently. If FIM $F(\boldsymbol{\theta})$ is decomposed as
\begin{equation}
    F(\boldsymbol{\theta}) = \sum_i \lambda_i \mathbf{v}_i \mathbf{v}_i^T,
\end{equation}
where $\lambda_i$ and $\mathbf{v}_i$ are eigenvalues and eigenvectors respectively, and $\mathbf{v}_i^T \mathbf{v}_j = \delta_{ij}$, one can consider the shifts $\boldsymbol{\theta}' = \boldsymbol{\theta} + \mathbf{v}_i \delta\theta$ for each separate direction $\mathbf{v}_i$. The direct analog of the scalar FI $F(\theta')$ for 1-parameter case is the projection of FIM for the shifted parameters at the considered direction:
\begin{equation}
    \lambda_i(\delta \theta) = \mathbf{v}_i^T F(\boldsymbol{\theta} + \mathbf{v}_i \delta\theta) \mathbf{v}_i.
\end{equation}
Therefore, the generalization of Eq.~(\ref{eq:regularized FIM 1D}) to the multi-parameter case is
\begin{equation}
    \tilde F (\boldsymbol{\theta}) = \sum_i \tilde\lambda_i \mathbf{v}_i \mathbf{v}_i^T
\end{equation}
where
\begin{equation}
    \tilde \lambda_i = \max_{\delta \theta} \frac{\lambda_i (\delta \theta)}{\left(1 + |\delta \theta| \sqrt{\lambda_i(\delta \theta)}\right)^2}.
\end{equation}

It is worth noting that we neither provide accurate proof of the proposed regularization procedure nor claim that the constructed modified FIM $\tilde F(\boldsymbol{\theta})$ ensures a rigorous bound via CRB for the estimation error. Rather, the approach is expected to be a useful practical tool for approximate assessment of the errors and resolution for multiparametric problems in quantum imaging. The next section illustrates the approach with several examples, starting from a simple 1-parameter toy example and covering resolution estimation in practical multiparametric quantum imaging.

\section{Application of the approach to quantum imaging}
\label{sec:applications}

\subsection{One-parameter model}
\label{subsec:1D}

To illustrate the proposed procedures of FIM correction for constrained problems in quantum imaging and assess the accuracy of the approach, we start from a simple toy example of a uniform object with the transmission amplitude $A$, $0 \le A \le 1$, representing the only parameter of interest. Let us assume that the light source emits groups of $n$ time-correlated photons and the $n$-photon coincidence events are detected. For $n = 2$, such process corresponds to emission of biphotons by spontaneous parametric down-conversion (SPDC) in a nonlinear crystal. Let the number of the emitted photon groups per an experiment have the mean value $N$ and be described by Poisson distribution. Then, the detected signal $Y$ has the expectation value
\begin{equation}
    S(A) = N \eta^n A^{2n},
\end{equation}
where $\eta$ is the light collection efficiency. 

The likelihood for the signal $Y$ (taking integer values) corresponds to Poisson distribution (\ref{eq:Poisson likelihood}) with the parameters vector $\boldsymbol{\theta}$ containing the only component $A$. Equation~(\ref{eq:FIM Poisson}) for FIM is applicable in the considered case and yields the result
\begin{equation}
    \label{eq:FIM 1D}
    F(A) = 4 n^2 N \eta^n A^{2(n-1)}.
\end{equation}

It is worth noting that the considered toy problem can be parameterized in different ways. For example, one can choose the probability of photon transmission $T = A^2$ as the parameter of interest instead of the transmission amplitude $A$ and get a different expression for the FI (Appendix~\ref{app:parameterization}). However, such parameterization would be quite unnatural for a more general multi-parameter case with light interference effects: for example, the transmission amplitude directly enters the expression (\ref{eq:G2 discretized}) for imaging with non-ideal spatial correlations of biphotons. The main goal of the current subsection is illustration of the developed approach by a simple example and preparation for analysis of practical multiparametric problems in Subsection~\ref{subsec:practical_applications}. For that reason, we follow the general course here and use $A$ as the parameter of interest, rather than search for the optimal parameterization of the specific toy problem. Further discussion of parameterization influence on FI is provided in Appendix~\ref{app:parameterization}.

As references for evaluation of the proposed approach, we consider the following two estimates of the transmission amplitude $A$:
\begin{itemize}
    \item Constrained MLE: 
    \begin{multline}
        \label{eq:MLE 1D definition}
        \hat A_\text{MLE}(Y) = \argmax_{0 \le A' \le 1} L(Y | S(A')) = \min \bigl(1, S^{-1}(Y)\bigr) \\ =  \min \bigl(1, Y^{1/(2n)} N^{-1/(2n)} \eta^{-1/2} \bigr).
    \end{multline}
    \item Mean over posterior probability distribution according to Bayes' formula:
    \begin{equation}
        \label{eq:Bayesian 1D definition}
        \hat A_\text{B}(Y) = \frac{\int_0^1 dA' L(Y|S(A')) A'}{\int_0^1 dA' L(Y|S(A'))}.
    \end{equation}
\end{itemize}

It is worth noting that the latter one also represents the optimal biased estimate (following the spirit of Refs.~\cite{benHaim2009lower,eldar2004minimum,eldar2006uniformly}), minimizing the mean squared error (MSE) averaged over the range of the true values $A \in [0, 1]$ (Appendix~\ref{app:optimal bias}).

\begin{figure}
    \centering
    \includegraphics[width=1.0\linewidth]{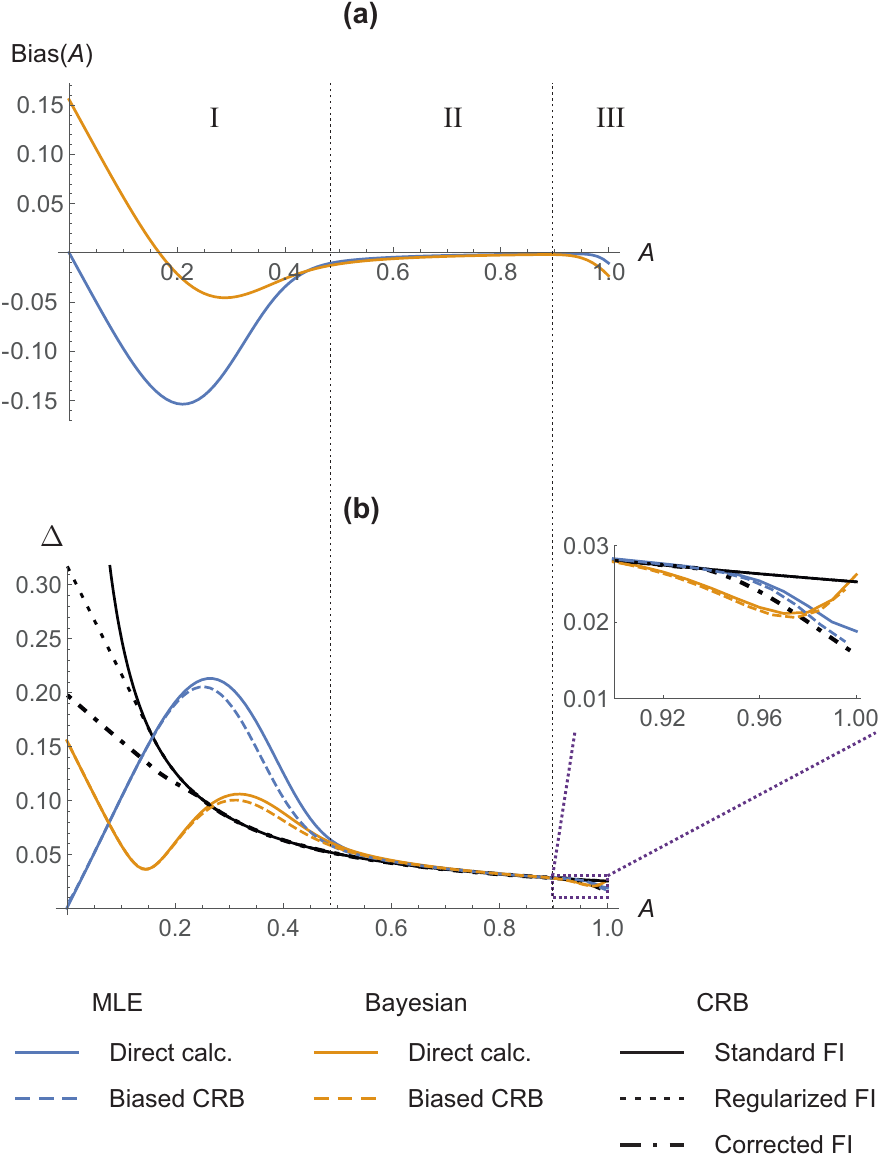}
    \caption{Transmission amplitude estimation for a uniform object (single-parameter model): estimation bias (a) and error (b). The notations of line styles are listed in the text. The regions I and III correspond to biased estimates for dark ($A$ close to 0) and bright ($A$ close to 1) objects respectively. In region II, the value $A$ is far from the boundaries and the estimate is almost unbiased. The inset in panel (b) shows a zoomed-in part of the region III, where the constraint $A \le 1$ leads to the estimation bias.}
    \label{fig:1D results}
\end{figure}

The bias of the estimates $\hat A_\text{MLE}$ and $\hat A_\text{B}$ is shown in Fig.~\ref{fig:1D results}(a) for $N =200$, $\eta = 0.7$, $n = 2$. One can see that the estimates are, indeed, biased near the boundary values $A = 0$ and $A = 1$ (in the regions I and III respectively).

Figure~\ref{fig:1D results}(b) shows the dependence of the estimation error
\begin{equation}
    \Delta = \sqrt{\operatorname{MSE}} = \sqrt{\operatorname{E}\bigl[(\hat A - A)^2\bigr]}
\end{equation}
on the true value $A$ of the transmission amplitude for the two estimators (blue and orange lines correspond to MLE and Bayesian estimators respectively). For each estimator, the MSE is calculated directly from the probability distribution (solid colored lines) and from modified CRB taking the bias into account \cite{eldar2004minimum} (dashed colored lines). The details of the calculations are described in Appendix~\ref{app:MSE for 1D}. 

One can notice that in the region I the error strongly depends on the estimator used; changing the bias, one can ``redistribute'' the inaccuracy over the range of the parameter $A$ \cite{eldar2008rethinking}. MLE is heavily biased to zero: it maps zero signal $Y = 0$ to $\hat A_\text{MLE}(0) = 0$, while such outcome remains the most probable one for the whole range $A \in [0, A_0]$ with $A_0 = N^{-1/(2n)} \eta^{-1/2} = 0.32$. The estimation error is vanishing for $A = 0$, but becomes quite large around the value $A = A_0$. On the opposite, the Bayesian estimate takes into account that zero signal is probable for $A \in [0, A_0]$ and maps it to $A_\text{B}(0) \approx A_0 \Gamma(1/n) / \Gamma(1/(2n)) = 0.49 A_0$. It is worth also mentioning the modified MLE, introduced in Appendix~5 of Ref. \cite{tsang2017subdiffraction}, where the MLE mapping is overridden by $A_\text{MLE}' = A_0 / 2$ to improve the worst-case estimation error. Still, the MSE remains finite for the whole range $A \in [0, 1]$ for the considered estimators. 

Black lines in Fig.~\ref{fig:1D results}(b) correspond to application of basic CRB (namely, Eq.~(\ref{eq:width by FIM 1D})) to the unmodified FI $F(A)$ (solid black line), FI after regularization according to Eq.~(\ref{eq:regularized FIM 1D}) (dotted black line; see Appendix~\ref{app:regularization 1D}), and FI $\tilde F(A)$ regularized and then corrected for the constraints (dot-dashed black line; see Appendix~\ref{app:correction 1D}). The regularization procedure resolves the divergence of the basic CRB with unmodified FI at $A = 0$. The correction, described in Subsection~\ref{subsec:effective FI}, further improves the error estimate and makes it consistent with the actual inaccuracies indicated by the colored lines in Fig.~\ref{fig:1D results}(b) (up to the strong variation of the error for different estimators in the region I).

\subsection{Two-parameter model}
\label{subsec:2D}

As the next step of the proposed procedure illustration, we consider the following 2-parameter model. The object consists of two parts with the transmission amplitudes $A_1$ and $A_2$, representing the parameters of interest: $\boldsymbol{\theta} \equiv \mathbf{A} =  (A_1, A_2)$. The physical domain of the parameters' values is naturally defined as $\Omega = \{ \mathbf{A} | 0 \le A_i \le 1 \}$. We assume that the object is illuminated by an SPDC light source, producing in average $N$ photon pairs ($n = 2$) per an experiment, and the second-order autocorrelations are measured by a pair of detectors with the efficiency $\eta$. The signal vector $\mathbf{S}(\mathbf{A})$ contains two components, which are described according to Eq.~(\ref{eq:G2 discretized}) as 
\begin{equation}
    S_i (\mathbf{A}) = N \eta^2 G^{(2)}_{ii} = N \eta^2 \biggl| \sum_{m,l=1,2} D^{(ii)}_{ml} A_m A_l \biggr|^2,\; i = 1,2.
\end{equation}

For simplicity, we assume that the coefficients $D^{(ii)}_{ml}$ are real, symmetric, and correspond to ideal spatial correlations of photons:
\begin{multline}
    D^{(ii)}_{ml} \in \mathcal{R},\; D^{(11)}_{11} = D^{(22)}_{22} = h_0, \\ D^{(11)}_{22} = D^{(22)}_{11} = h_1,\;  D^{(ii)}_{12} = D^{(ii)}_{21} = 0.
\end{multline}
Therefore, the signal components can be expressed as
\begin{equation}
    S_1(\mathbf{A}) = N \eta^2 \left( h_0 A_1^2 + h_1 A_2^2\right)^2,
\end{equation}
\begin{equation}
    S_2(\mathbf{A}) = N \eta^2 \left( h_1 A_1^2 + h_0 A_2^2\right)^2.
\end{equation}
The realizations of the measured signal $\mathbf{Y} = (Y_1, Y_2)$ are assumed to be characterized by Poisson distribution with $Y_i$ having the expectation value $S_i(\mathbf{A})$. The FIM for the problem can be calculated according to Eq.~(\ref{eq:FIM Poisson}) and equals
\begin{equation}
    \label{eq:FIM 2D}
    F (\mathbf{A}) = 16 N \eta^2 \left(h_0^2 + h_1^2 \right) \left(
    \begin{array}{cc}
        A_1^2 & \zeta A_1 A_2 \\
        \zeta A_1 A_2 & A_2^2
    \end{array}
    \right),
\end{equation}
where
\begin{equation}
    \zeta = \frac{2 h_0 h_1}{h_0^2 + h_1^2}, \quad |\zeta| <1 \text{ for } h_0 \ne h_1.
\end{equation}

For comparison with the error prediction by the proposed approach, we consider the following  estimates of the parameters of interest $\mathbf{A}$:
\begin{itemize}
    \item Constrained MLE:
    \begin{equation}
        \hat{\mathbf{A}}_\text{MLE}(\mathbf{Y})=\argmax_{\mathbf{A}' \in \Omega} \left[ L(\mathbf{Y} | \mathbf{S}(\mathbf{A}')) \right]
    \end{equation}
    where the likelihood for the 2-component signal is defined according to Eq.~(\ref{eq:Poisson multivariate likelihood}).
    \item Bayesian mean \textit{a posteriori} estimate:
    \begin{equation}
        \hat{\mathbf{A}}_\text{B}(\mathbf{Y}) = \frac{\int_0^1 dA_1'\int_0^1 dA_2' L(\mathbf{Y}|\mathbf{S}(\mathbf{A}')) \mathbf{A}'}{\int_0^1 dA_1'\int_0^1 dA_2' L(\mathbf{Y}|\mathbf{S}(\mathbf{A}'))}.
    \end{equation}
\end{itemize}

\begin{figure*}
    \centering
    \includegraphics[width=0.9\linewidth]{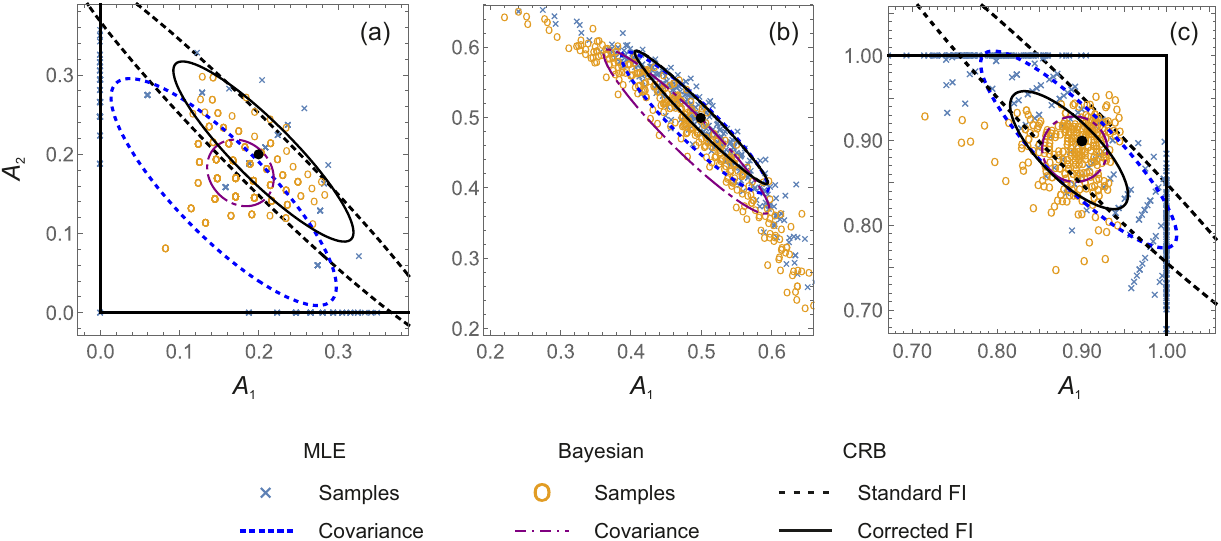}
    \caption{Transmission amplitudes estimation for the 2-parameter model. The true parameters values are $A_1 = A_2 = 0.2$ (a), 0.5 (b), and 0.9 (c), as indicated by black dots. Blue crosses and orange circles show the parameter inference results of a randomly sampled signal for MLE and Bayesian estimate respectively. 1000 samples were used for statistics estimation; 300 samples are shown in the figure; some results coincide due to discretness of the signal. The modeling parameters are $\eta = 0.7$, $h_0 = 1$, $h_1 = 0.8$; $N = 1000$ (a,b) and 50 (c). Ellipses (blue dotted for MLE and purple dot-dashed for Bayesian estimate) are centered at the mean of the sampled results and have the shape defined by the corresponding covariance matrix (Appendix~\ref{app:visualization 2D}). Black dotted ellipse (extending beyond the plotted regions) shows the predicted results scattering according to CRB with the standard FIM $F(\mathbf{A})$. Black solid ellipse corresponds to the regularized and corrected FIM $\tilde F (\mathbf{A})$. Straight black lines show the constraints imposed on the parameters $A_1$ and $A_2$.}
    \label{fig:2D results}
\end{figure*}

Figure~\ref{fig:2D results} shows the results for the considered 2-parameter model. The estimators listed above are applied to 1000 random samples of the signal $\mathbf{Y}$ for the true values of the transmission amplitudes localized in the lower part (a), the center (b), and the upper part (c) of the physical domain. The estimation results $\mathbf{Y} \mapsto \hat{\mathbf{A}}$ are shown as points in each part of the figure. The statistics of each generated dataset is shown as an ellipse, indicating the mean value $\langle \hat{\mathbf{A}} \rangle$ and the covariance $C = \operatorname{Cov}(\hat{\mathbf{A}},\hat{\mathbf{A}})$ according to the approach described in Appendix~\ref{app:visualization 2D}:
\begin{equation}
    \Delta \mathbf{A}^T C^{-1} \Delta \mathbf{A} = 2 \log 2,\quad \Delta \mathbf{A} = \mathbf{A} - \langle \hat{\mathbf{A}} \rangle.
\end{equation}

The results of random sampling are compared with the error predictions based on CRB and FIM: the standard one $F(\mathbf{A})$, calculated according to Eq.~(\ref{eq:FIM 2D}), and the FIM $\tilde F(\mathbf{A})$ obtained by applying the proposed regularization and correction procedures to $F(\mathbf{A})$. The FIMs $F(\mathbf{A})$ and $\tilde F(\mathbf{A})$ are also visualized as ellipses (Appendix~\ref{app:visualization 2D}).

First, one can notice that in the case, shown in Fig.~\ref{fig:2D results}(b), the true value $\mathbf{A} = (0.5, 0.5)$ is far from the boundaries. The regularization and correction procedures act trivially: $\tilde F(\mathbf{A}) = F (\mathbf{A})$ and the dashed and solid black ellipses coincide. The statistics of MLE and Bayesian estimates almost coincide with each other and with the prediction according to CRB (still, some estimation bias is present).

For the cases when the constraints are active (Fig.~\ref{fig:2D results}(a,c)), CRB with standard FIM strongly overestimates the errors. The sampling statistics depends on the particular estimator used. Up to that variability, the corrected FIM provides a reasonable prediction of the parameters estimation inaccuracies.

\subsection{Practical multiparametric models in quantum imaging}
\label{subsec:practical_applications}

\begin{figure}
    \centering
    \includegraphics[width=0.85\linewidth]{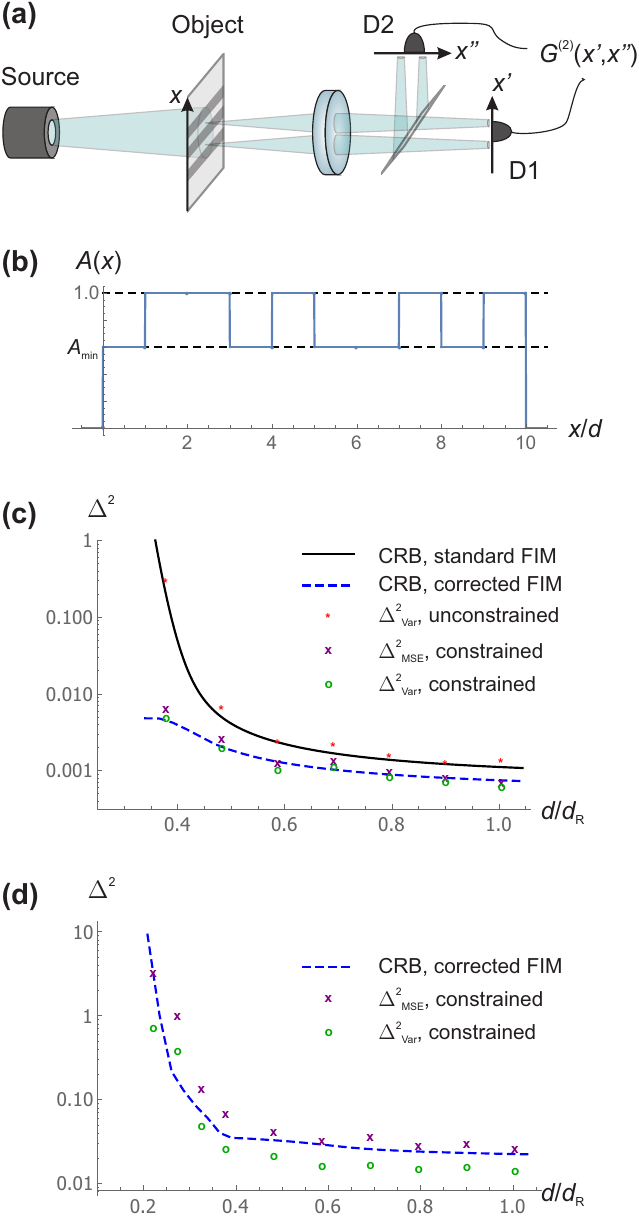}
    \caption{Transmission amplitudes estimation for the 10-parameter model. (a) Scheme of the modeled imaging setup. Biphoton radiation, passing through the investigated object, is collected by an optical system and split between two single-photon detectors. The signal is represented by the correlation function $G^{(2)}(x',x'')$ depending on the detectors' positions $x'$ and $x''$. (b) Transmission amplitude of the model object as a function of the transverse coordinate $x$ normalized by the width $d$ of the slits (extended pixels) used for pixeliation of the object. (c,d) The dependence of the estimation error on the slit width $d$ normalized by the Rayleigh resolution limit $d_\text{R}$ for $A_\text{min} = 0.9$ (c) and $A_\text{min} = 0$ (d). The notations and line styles are described in the text.}
    \label{fig:multiparameter simple results}
\end{figure}

To conclude testing of the proposed approach, we apply it to multiparametric quantum imaging problems analyzed in Refs.~\cite{mikhalychev2019efficiently,mikhalychev2021lost}. The considered imaging setup is schematically shown in Fig.~\ref{fig:multiparameter simple results}(a) and follows the ideas of the biphoton imaging experiment in Ref.~\cite{mikhalychev2019efficiently}. An object consists of vertical slits (each having width $d$ and a constant transmission amplitude --- Fig.~\ref{fig:multiparameter simple results}(b)) and is illuminated by entangled photon pairs. The radiation is collected by an optical system with the resolution limit $d_\text{R}$ (classical Rayleigh limit for incoherent imaging). The registered signal represents coincidence counts number for a pair of detectors.

First, we consider a 10-parameter model with the object shown in Fig.~\ref{fig:multiparameter simple results}(b) \cite{mikhalychev2021lost} and parameterized by the transmission amplitudes $A_m$ (each describing the region $(m-1)d \le x \le md$ of the object): $\boldsymbol{\theta} \equiv \mathbf{A} = (A_1,\ldots, A_{10})$. We choose $A_\text{min} = 0.9$ for testing the correction-only procedure and $A_\text{min} = 0$ as the case when both regularization and correction of FIM are required. 

The incident photon pairs are assumed to be ideally correlated, and the signal is described by Eq.~(\ref{eq:G2 discretized}) with $i = j$, $m = l$, and the sampling points corresponding to the step $d / 2$ in the object plane:
\begin{equation}
    \label{eq:signal lost-photon model}
    S_j(\mathbf{A}) = N G^{(2)}(x_j, x_j) = N \biggl| \sum_m D^{(jj)}_{mm} A_m^2 \biggr|^2,
\end{equation}
where $N$ is the expectation value of the total number of detected coincidence events (already taking into account the detectors efficiency), $x_j = jd / 2$,
\begin{equation}
    D^{(jj)}_{mm} = 4 k_\text{max}^2 \int_{(m-1)d}^{md} ds \operatorname{sinc}^2 \left[k_\text{max}\left(s - x_j \right)\right],
\end{equation}
and $k_\text{max} = 3.83 / d_\text{R}$ is the momentum transfer cut-off of the optical system \cite{mikhalychev2021lost}.

Similarly to previous examples, Poisson statistics of the signal is assumed, and the FIM is calculated according to Eq.~(\ref{eq:FIM Poisson}). The total value of the estimation error, predicted by CRB, can be quantified by the quantity $\Delta^2$ defined in Eq.~(\ref{eq:Delta_squared_def}) and applied to both the standard FIM $F(\mathbf{A})$ and its regularized and corrected counterpart $\tilde F(\mathbf{A})$. As a reference for quality assessment of the predictions, we also perform direct estimation of the transmission amplitudes $\hat{\mathbf{A}}$ for the signal $\mathbf{Y}$ randomly sampled according to its likelihood $L(\mathbf{Y}|\mathbf{S}(\mathbf{A}))$. The inference is performed by minimizing the distance between the realized signal $\mathbf{Y}$ and its parameterized expectation value $\mathbf{S}(\hat{\mathbf{A}})$:
\begin{equation}
    \label{eq:multiparametric estimation}
    \hat{\mathbf{A}} = \argmin_{\mathbf{A'}} \left| \mathbf{Y} - \mathbf{S}(\mathbf{A'})  \right|^2.
\end{equation}
The estimation inaccuracy is characterized by the total variance and MSE for the sampling results:
\begin{equation}
    \Delta^2_\text{Var} = \sum_m \operatorname{Var}(\hat{A}_m),\quad \Delta^2_\text{MSE} = \bigl\langle \bigl| \hat{\mathbf{A}} - \mathbf{A} \bigr|^2 \bigr\rangle.
\end{equation}

Figure~\ref{fig:multiparameter simple results}(c) shows the dependence of the estimation error $\Delta^2$ on the spatial scale of the problem $d / d_\text{R}$ for the model object presented in Fig~\ref{fig:multiparameter simple results}(b) with $A_\text{min} = 0.9$. The number of the detected coincidence events is taken to be $N = 10^4$. The error $\Delta^2$ calculated for the standard FIM $F(\mathbf{A})$ (solid line) accurately describes the variance for unconstrained estimation (red star-marked points). If the constraint $0 \le A_m \le 1$ is imposed during the transmission amplitudes inference according to Eq.~(\ref{eq:multiparametric estimation}), the variance (green circles) and MSE (purple crosses) become smaller. Such error decrease is accurately reproduced by using CRB with the corrected FIM $\tilde F(\mathbf{A})$ (dashed blue line).

Figure~\ref{fig:multiparameter simple results}(d) presents similar results for the case $A_\text{min} = 0$, which requires regularization of FIM. Application of CRB to the standard FIM does not yield meaningful results in that case since $F(\mathbf{A})$ is singular. For that reason, only the constrained estimation was sampled and compared to the error prediction by the regularized and corrected FIM $\tilde F(\mathbf{A})$. One can see that the agreement between the results of the proposed approach and direct sampling is good.

As the last example, we consider the object from a model quantum imaging experiment presented in Ref.~\cite{mikhalychev2019efficiently}. The spatial resolution, achieved in practice, was higher than the theoretical bound predicted by CRB with the standard FIM. Such counter-intuitive result in Ref.~\cite{mikhalychev2019efficiently} was the motivation for the current research. The measurement setup in the discussed experiment corresponded to the scheme in Fig.~\ref{fig:multiparameter simple results}(a). The resolution of the optical system was artificially worsened by decreasing the numerical aperture for demonstration of sub-Rayleigh imaging and corresponded to the Rayleigh limit $d_\text{R} = 30.0$~$\mu$m. In contrast to the previous idealized 10-parameter imaging model, the spatial correlation of biphotons is not assumed to be ideal here (i.e., the photons from a pair have non-zero probability of passing through different parts of the object). 

\begin{figure}
    \centering
    \includegraphics[width=0.97\linewidth]{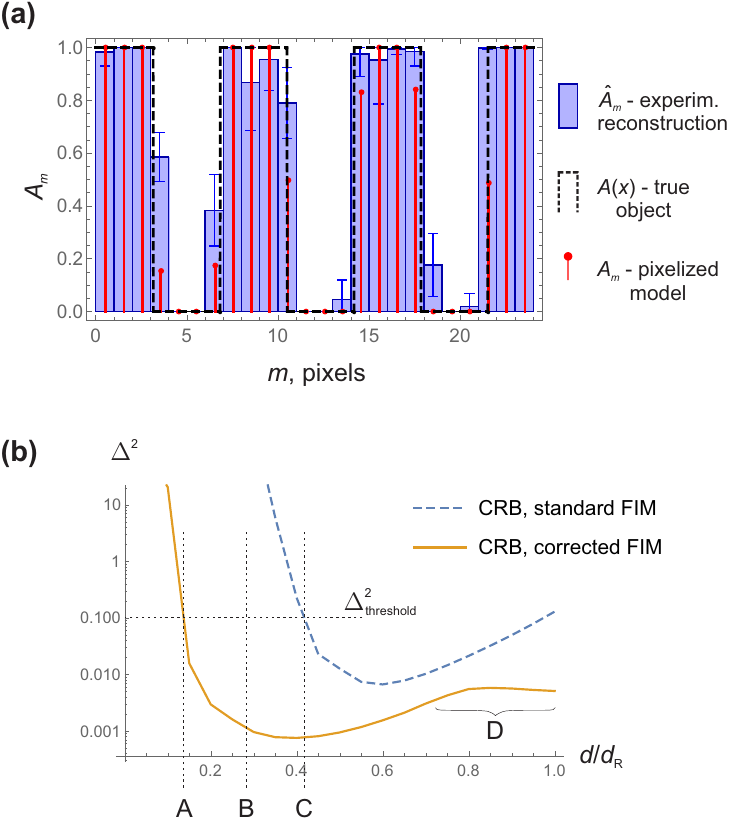}
    \caption{Transmission amplitudes estimation for the 24-parameter model from Ref.~\cite{mikhalychev2019efficiently}. (a) Transmission amplitude of the model object: pixelized reconstruction result $\hat{\mathbf{A}}=\{\hat A_m\}$ from Ref.~\cite{mikhalychev2019efficiently} (filled blue rectangles) with error bars built on the base of repeated measurement results; true transmission amplitude of the model object $A(x)$ (dashed black line); and its pixelized counterpart representing the true values $\mathbf{A}= \{A_m\}$ of the parameters of interest for the solved problem (red vertical bars). (b) The dependence of the estimation error on the slit width $d$ normalized by the Rayleigh resolution limit $d_\text{R}$ via CRB with the standard FIM (dashed blue line) and the regularized and corrected FIM (solid orange line). The dotted vertical lines indicate characteristic resolution values: A --- the resolution limit for the corrected FIM, defined for the threshold error value $\Delta^2_\text{threshold} = 0.1$; B --- the spatial scale of the successful experimental inference of the object transmission amplitude \cite{mikhalychev2019efficiently}; C --- the resolution limit for the standard FIM. In region D, the regularizing effect of the non-ideal biphoton correlations decreases with the growth of the problem scale $d$.}
    \label{fig:Supertwin results}
\end{figure}

The object is shown in Fig.~\ref{fig:Supertwin results}(a). Physically, it represents 3 non-transparent (dark) slits on a transparent background (a part of a positive USAF 1951 resolution target) with the slit width 31.25~$\mu$m. For statement of the estimation problem, the object was divided into 24 slit-like pixels and parameterized by 24 values of the transmission amplitude $\boldsymbol{\theta} \equiv \mathbf{A} = (A_1, \ldots, A_{24})$. The true values $\mathbf{A}$ used for our modeling (Fig.~\ref{fig:Supertwin results}(a)) corresponded to the actual object used in the experiment \cite{mikhalychev2019efficiently}.

The measured signal $\mathbf{S}(\mathbf{A})$ is represented by the full correlation matrix $G^{(2)}(x_i, x_j)$ without the constraint $i = j$ \cite{mikhalychev2019efficiently}: 
\begin{equation}
    \label{eq:signal general model}
    S_{ij}(\mathbf{A}) = N G^{(2)}(x_i, x_j) 
\end{equation}
where $G^{(2)}(x_i, x_j)$ is calculated according to Eq.~(\ref{eq:G2 discretized}),  the expressions for the coefficients $D^{(ij)}_{ml}$ are provided in Ref.~\cite{mikhalychev2019efficiently}, and $N$ is the expectation value of the total number of detected coincidence events. As previously, we assume independent Poissonian statistics for the components $Y_{ij}$ of the measured signal $\mathbf{Y}$. Therefore, FIM is described by Eq.~(\ref{eq:FIM Poisson}) and takes the form
\begin{equation}
    \label{eq:FIM Supertwin}
    F_{ml}(\mathbf{A}) = 4 N \sum_{i,j} \Bigl( \sum_{l'} D_{ij}^{(ml')} A_{l'}\Bigr) \Bigl( \sum_{m'} D_{ij}^{(m'l)} A_{m'}\Bigr).
\end{equation}
Typically, it is sufficient to have at least one non-dark pixel ($\exists l: A_l \ne 0$) to ensure non-zero values of the expressions in parentheses and, therefore, non-singular FIM $F(\mathbf{A})$. The non-ideal biphoton correlations produce regularizing effect by shifting the problem closer to coherent light imaging, where the FIM singularity issue does not arise.

Figure~\ref{fig:Supertwin results}(b) shows the results of resolution estimation via CRB for $N = 10^5$. The total reconstruction error $\Delta^2$, defined by Eq.~(\ref{eq:Delta_squared_def}), is calculated for the standard FIM $F(\mathbf{A})$ (dashed blue line) and the regularized and corrected one $\tilde F(\mathbf{A})$ according to the proposed approach (solid orange line). The achievable resolution can be defined as the minimal problem scale (pixel size) $d = d_\text{min}$, for which the reconstruction error $\Delta^2$ remains within certain reasonable limit: $\Delta^2 \le \Delta^2_\text{threshold}$ \cite{mikhalychev2021lost}. For illustration purposes, the threshold value $\Delta^2_\text{threshold} = 0.1$, consistent with the actual experimental noise in Ref.~\cite{mikhalychev2019efficiently}, is used in Fig.~\ref{fig:Supertwin results}(b). For such selection of the threshold, the theoretical resolution limit for the standard FIM is $d_\text{min} / d_\text{R} \approx 0.42$, which is larger than the spatial scale $d / d_\text{R} = 8.51\text{ $\mu$m} / 30.0\text{ $\mu$m} = 0.28$ of the problem, successfully solved experimentally \cite{mikhalychev2019efficiently}. FIM correction according to the proposed approach shifts the resolution limit to the value $d_\text{min} / d_\text{R} \approx 0.14$, thus, resolving the seeming contradiction of beating the ultimate theoretical limit experimentally. 

An interesting and rather counter-intuitive feature of the obtained dependencies is the growth of the predicted error $\Delta^2$ when the spacial scale $d$ of the problem increases (region D). The effect is closely related to the regularizing influence of the finite transverse correlation length of non-ideal biphotons, discussed after Eq.~(\ref{eq:FIM Supertwin}). When the pixel size increases and the correlation length remains constant, the cross-term coefficients $D_{ml}^{(ij)}$ with $m \ne l$ tend to zero and the standard FIM becomes ill-defined. However, the effect is less pronounced for the regularized and corrected version of FIM.

\section{Conclusions}

\textit{A priori} constraints, imposed on quantum imaging problems, can improve stability of the object parameters inference and decrease the errors beyond the CRB-based theoretical limit (strictly speaking, inapplicable due to estimation bias). The approach, proposed in the current paper, restores the applicability of the usual form CRB to such constrained problems by constructing a modified FIM, which includes the constraints and is regularized if the initial (standard) FIM turns out to be ill-defined.

The proposed FIM correction and regularization technique represents a practical tool suitable for complex multiparametric models and accurate enough to be used for resolution analysis. A competitive approach, based on modified CRB \cite{eldar2004minimum,eldar2006uniformly}, requires knowledge of the bias derivatives, which may be infeasible for analytical calculation and very time-consuming for estimation by numerical Monte-Carlo simulations. Even for a relatively simple 2-parameter case, the analytical description of biased constrained estimates is cumbersome.

The proposed approach was successfully applied to 1-, 2-, 10-, and 24-parameter imaging problems and exhibited a good agreement with other methods of error estimation. For the simplest 1-parameter model, it is possible to describe various estimators and to quantify resolution analytically. Many fundamental theoretical works have already been devoted to such simple models. For 2 parameters, it is still possible to estimate the errors directly, but the calculations are more complicated. When a multiparametric problem is considered, the proposed procedure continues to work well and yields meaningful results, while other approaches require huge computational efforts. In particular, the developed technique of FIM modification helped us to explain the high resolution achieved experimentally in Ref.~\cite{mikhalychev2019efficiently} and to predict the resolution limit for such a constrained problem quantitatively. While the provided illustrations assumed Poisson distribution of the signal noise and applicability of Eq.~(\ref{eq:FIM Poisson}), the proposed method itself is not limited to such models in any way and can be applied to FIM regardless of the specific type of expression used for its calculation. It is worth noting that the applicability of the developed approach can be extended to problems with a larger number of parameters (e.g., hundreds or thousands) by analyzing sub-problems of a reasonable size after splitting the initial problem into almost independent windows \cite{mikhalychev2019efficiently}.

The ability to construct the modified FIM for constrained problems can be useful for resolution estimation (as shown in the last considered example), problem structure analysis (similar to the one performed in Ref.~\cite{mikhalychev2019efficiently}, but more accurate for constrained problems), imaging setup optimization, and other similar tasks. The applicability of the approach extends far beyond quantum optics. For example, Monte-Carlo simulations for high-resolution X-ray diffraction in Ref.~\cite{mikhalychev2021fisher} demonstrated decrease of the estimation variance when the natural physical constraints imposed on the parameters of interest became active. The proposed technique is capable of predicting such effect on the base of the modified FIM and CRB.

Since QFI equals the classical FI for the optimal measurement, it is affected by the \textit{a priori} constraints in a very similar way. Therefore, the developed approach is not limited to classical FI, and can be applied to QFI as well.


\begin{acknowledgments} 
A.M., S.M., and D.M. acknowledge support from the National Academy of Sciences of Belarus program ``Convergence''.
A.M. and S.M. acknowledge partial support from BRFFR grant F24KI-058 and from MSHE-RF grant 075-15-2024-556. 
A.M. and D.M. also acknowledge partial support from BRFFR grant F23UZB-064.
\end{acknowledgments}
\appendix

\section{Geometry of likelihood and Fisher information}
\label{app:geometry of FI}

\begin{figure}
    \centering
    \includegraphics[width=0.8\linewidth]{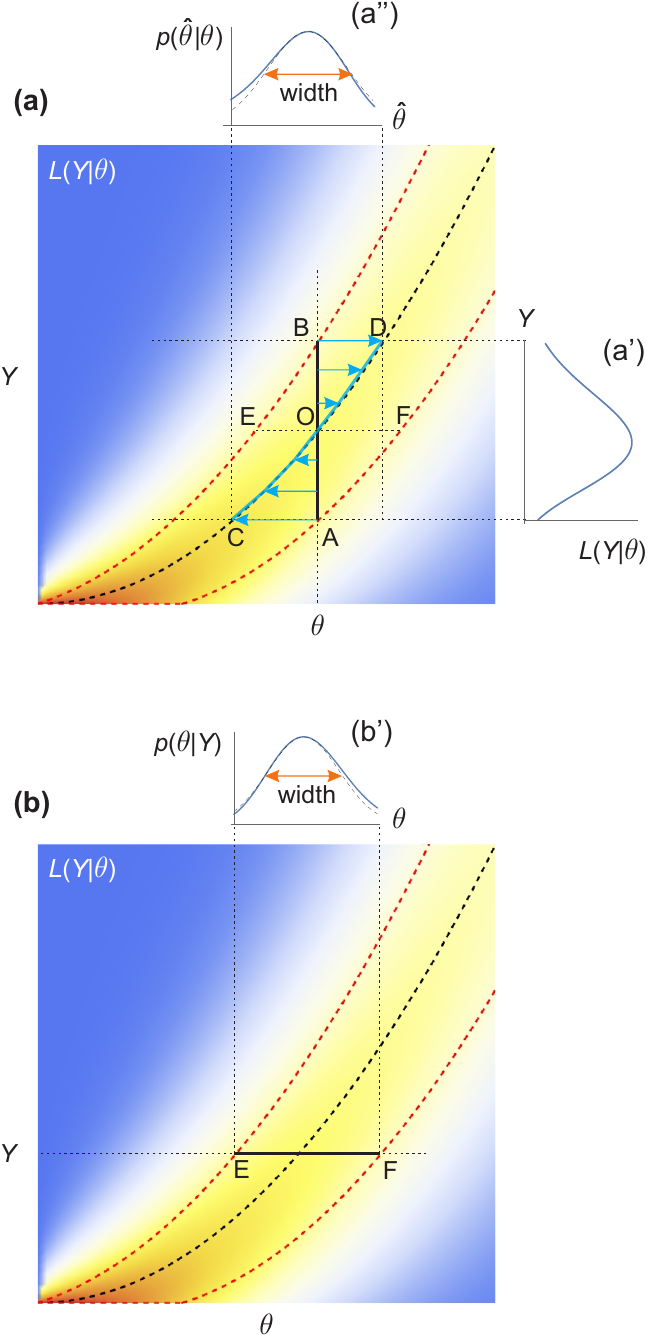}
    \caption{Geometry of likelihood and FI: (a) distribution of the estimated parameter value $\hat \theta$ for a given true value $\theta$ of the parameter; (b) Bayesian posterior probability for the parameter $\theta$ for a given signal value $Y$. The details are provided in the text of Appendix~\ref{app:geometry of FI}. The segment EF is also shown in panel (a) for the purpose of comparing the two interpretations. For simplicity, the signal $Y$ is shown as a continuous variable, while for quantum-optical measurements it is typically a discrete variable taking integer values (the number of detected photons or coincidence events).}
    \label{fig:Fisher geometry}
\end{figure}

The close connection of FI with distribution of the estimation results (via CRB) on the one hand and with the likelihood (via the definition (\ref{eq:FIM_score})) on the other hand gives rise to its frequentist and Bayesian interpretations (see e.g. Refs. \cite{jaynes1996,vallisneri2008}). Geometry of the connection between the likelihood, distribution of the estimation results, and FI is illustrated by Fig.~\ref{fig:Fisher geometry}. The maps schematically (without direct relation to any specific physical model) represent the likelihood $L(Y | \theta)$ for a simple 1-parameter problem with a scalar (1-dimensional) signal $Y$. 

Figure~\ref{fig:Fisher geometry}(a) corresponds to interpretation of FI in terms of the estimation results distribution for a fixed model. If the true value of the parameter equals $\theta$, the possible realizations of the detected signal $Y$ are distributed according to the likelihood $L(Y | \theta)$, shown in the inset (a') for specific $\theta$ and representing the vertical cross section AB of the map. Inference of the investigated parameter through its MLE $\hat \theta$ corresponds to processing the detected signal value $Y$ by mapping $Y \mapsto \hat \theta = \argmax_{\hat \theta} L(Y |\hat \theta)$. Geometrically, each point of the segment AB is mapped onto the curve CD.  The resulting distribution $p(\hat \theta | \theta)$, shown in the inset (a''), describes frequencies of getting the estimated values $\hat \theta$ in a detection-inference procedure if the true value of the investigated parameter equals $\theta$. In the high-SNR limit, the distribution can be approximated by a Gaussian profile (dashed line in inset (a'')). The width of the distribution (total length $l(\text{CA}) + l(\text{BD})$ of the segments CA and BD) characterizes the estimation error and is determined by FI.

The geometry for the Bayesian interpretation of FI is illustrated in Fig.~\ref{fig:Fisher geometry}(b). For a given value of the measured signal $Y$, the posterior distribution $p(\theta|Y)$ of the parameter $\theta$  (inset (b')) represents a horizontal cross-section EF of the likelihood map. In the high-SNR limit, the Gaussian approximation of the posterior probability is valid (dashed line in the inset (b')), with it width $l(\text{EF})$ determined by the FI according to Eq.~(\ref{eq:FIM-posterior}).

By overlaying the segment EF onto the plot in Fig.~\ref{fig:Fisher geometry}(a), one can see that, in the high-SNR limit, the two interpretations of the FI lead to the same predicted error (width of the probability distribution). In that limit, the curve CD is almost a straight line. Therefore, the shapes EOB and OBD, as well as COA and AOF are close to triangles, which are equal to each other pairwise. The width $l(\text{EF})$ of the posterior probability distribution in Bayesian interpretation is approximately the same as the deviation between the estimated and the true values of the parameter in the frequentist approach:
\begin{equation}
    l(\text{EF}) = l(\text{EO}) + l(\text{OF}) \approx l(\text{BD}) + l(\text{CA}).
\end{equation}

\begin{figure}
    \centering
    \includegraphics[width=0.8\linewidth]{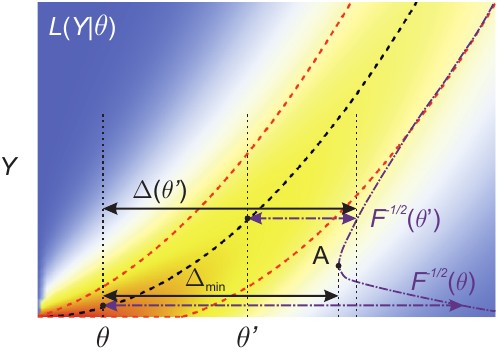}
    \caption{FI regularization for 1-parameter problem. The same likelihood diagram as in Fig.~\ref{fig:Fisher geometry} is used. Purple dot-dashed line schematically shows the error estimate (upper bound for $\hat \theta$) based on FI (Eq.~(\ref{eq:width by FIM 1D})), which diverges at $\theta \rightarrow 0$. The modified error estimate $\Delta(\theta')$ defined by Eq.~(\ref{eq:Delta theta shifted}) takes into account FI-based error estimate and the displacement $\theta' - \theta$ for a shifted parameter value $\theta'$. The resulting regularized error estimate $\Delta_\text{min}$ for $\theta$ is constructed by minimizing the quantity $\Delta(\theta')$ over $\theta'$ (Eq.~(\ref{eq:Delta min}) --- i.e., by finding the rightmost point A along the purple dot-dashed curve).}
    \label{fig:regularization geometry}
\end{figure}

The introduced geometrical treatment of the likelihood and FI is also useful for illustration of the regularization procedure described by Eqs.~(\ref{eq:Delta theta shifted}) and (\ref{eq:Delta min}) in Subsection~\ref{subsec:regularization} --- Fig.~\ref{fig:regularization geometry}.

\section{Fisher information for independent events}
\label{app:Poisson FIM}

For the specific case of the likelihood, defined by Eq.~(\ref{eq:Poisson multivariate likelihood}), the general expression~(\ref{eq:FIM_sum}) can be simplified in the following way. Due to additivity of FI \cite{lehmann1998} and independence of the signal components (factorization of the expression (\ref{eq:Poisson multivariate likelihood}) for the likelihood), the FIM is a sum of separate contributions $F_{\mu\nu}^{(i)}$:
\begin{equation}
    \label{eq:app:FIM_additivity}
    F_{\mu\nu} = \sum_i F_{\mu \nu}^{(i)}
\end{equation}
where a term
\begin{equation}
    \label{eq:app:FIM_single-component}
    F_{\mu \nu}^{(i)} = \sum_{Y_i = 0}^\infty \frac{1}{L(Y_i|S_i(\boldsymbol{\theta}))} \frac{\partial L(Y_i|S_i(\boldsymbol{\theta}))}{\partial \theta_\mu} \frac{\partial L(Y_i|S_i(\boldsymbol{\theta}))}{\partial \theta_\nu}
\end{equation}
represents the FI for the $i$-th component of the signal.

For the single-component likelihood, described by Eq.~(\ref{eq:Poisson likelihood}), the derivatives over the parameters can be expressed as
\begin{equation}
    \label{eq:app:Poisson derivative}
    \frac{\partial L(Y_i|S_i(\boldsymbol{\theta}))}{\partial \theta_\mu} = \frac{\partial S_i(\boldsymbol{\theta})}{\partial \theta_\mu} \left( \frac{Y_i}{S_i(\boldsymbol{\theta})} - 1 \right) L(Y_i|S_i(\boldsymbol{\theta})).
\end{equation}
Substitution of Eq.~(\ref{eq:app:Poisson derivative}) into Eq.~(\ref{eq:app:FIM_single-component}) yields the resulting expression
\begin{multline}
    \label{eq:app:FIM_single-component result}
    F_{\mu \nu}^{(i)} = \frac{\partial S_i(\boldsymbol{\theta})}{\partial \theta_\mu} \frac{\partial S_i(\boldsymbol{\theta})}{\partial \theta_\nu} \sum_{Y_i = 0}^\infty \left( \frac{Y_i}{S_i(\boldsymbol{\theta})} - 1 \right)^2 L(Y_i|S_i(\boldsymbol{\theta})) \\ = \frac{1}{S_i(\boldsymbol{\theta})} \frac{\partial S_i(\boldsymbol{\theta})}{\partial \theta_\mu} \frac{\partial S_i(\boldsymbol{\theta})}{\partial \theta_\nu}.
\end{multline}
Eq.~(\ref{eq:FIM Poisson}) is obtained by substitution of  expression~(\ref{eq:app:FIM_single-component result}) into Eq.~(\ref{eq:app:FIM_additivity}).

The presented approach to construction of FIM is based on treating the measured dataset as a whole and considering the multivariate probability distribution for the signal vector $\mathbf{Y}$. An alternative approach, typical for quantum imaging \cite{zhou2019quantum,tsang2016quantum,tsang2017subdiffraction,paur2018tempering}, is based on calculation of FI $F^\text{(event)}$ for a single detection event with subsequent multiplication of the result by the number of detected events $N$. A single detection event is characterized by the probability distribution $p_i (\boldsymbol{\theta})$, where the index $i$ enumerates the possible outcomes. For example, when just field intensity is measured ($n = 1$ in Eq.~(\ref{eq:Gn})), the index $i$ describes the position of a single-photon detection $\mathbf{r}^{(i)}$. The probabilities are normalized as $\sum_i p_i (\boldsymbol{\theta}) = 1$, and the FIM can be constructed according to Eq.~(\ref{eq:FIM_sum}) as
\begin{equation}
\label{eq:app:FIM probability-based}
    F_{\mu\nu} = N F_{\mu\nu}^\text{(event)} = N \sum_i \frac{1}{p_i(\boldsymbol{\theta})} \frac{\partial p_i(\boldsymbol{\theta})}{\partial \theta_\mu} \frac{\partial p_i(\boldsymbol{\theta})}{\partial \theta_\nu}.
\end{equation}

One can easily relate the single-event probability distribution $p_i(\boldsymbol{\theta})$ with the mean number $S_i(\boldsymbol{\theta})$  of such events used in our approach as
\begin{equation}
    S_i(\boldsymbol{\theta}) = N p_i(\boldsymbol{\theta}).
\end{equation}
Therefore, Eqs.~(\ref{eq:FIM Poisson}) and (\ref{eq:app:FIM probability-based}) yield exactly the same result. Such equivalence of the approaches is not surprising since the Poisson distribution assumed during derivation of Eq.~(\ref{eq:FIM Poisson}) is directly related to the independence of the events used in Eq.~(\ref{eq:app:FIM probability-based}).

\section{Iterative algorithm of Fisher information matrix modification}
\label{app:Iterative algorithm}

\begin{figure*}
    \centering
    \includegraphics[width=1.0\linewidth]{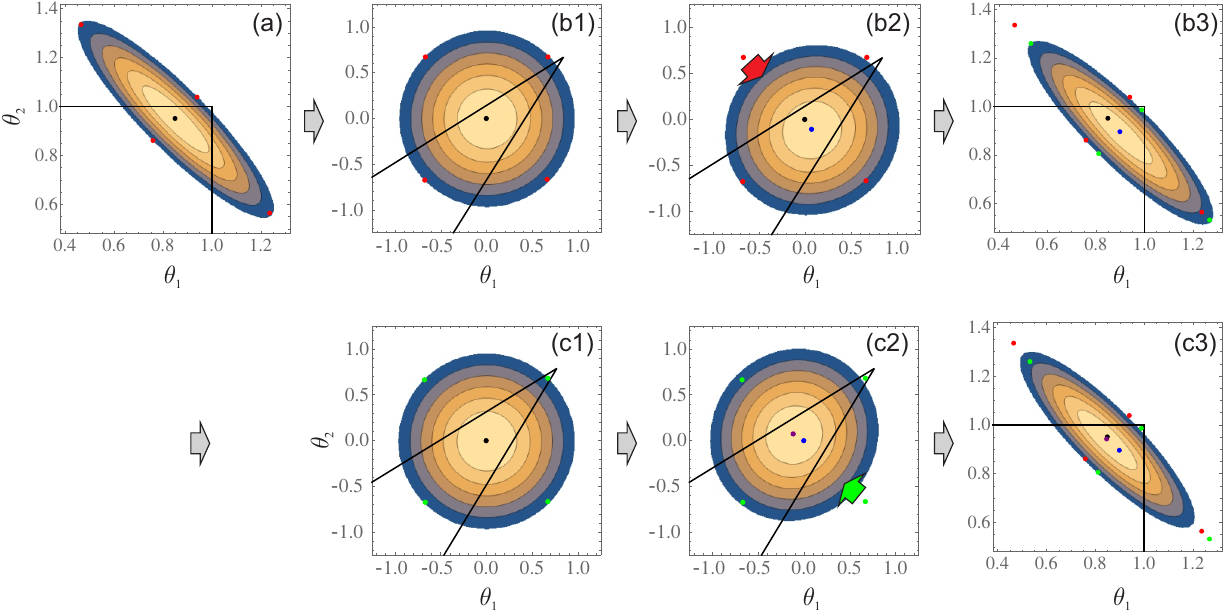}
    \caption{Iterative shrinking of a model probability distribution in 2-parameter case: initial distribution (a) and two subsequent iterations (b1--b3) and (c1--c3). The constraints are $\theta_1 \le 1$ and $\theta_2 \le 1$. The black dot indicates the true values of the parameters $(\theta_1, \theta_2) = (0.85, 0.95)$. Each iteration starts from the coordinate transformation described in the Subsection~\ref{subsec:effective FI} of the main text (b1, c1). Then, the resulting distribution is modified to reduce the probability of violating one of the constraints (b2, c2). Finally, the inverse coordinate transform if applied (b3, c3). Red and green points indicate the shape of the probability distribution before the first and the second iterations respectively. Blue and purple points show the mean values for the modified distributions after the first and the second iterations respectively.}
    \label{fig:iterations}
\end{figure*}

The iterative procedure of shrinking the multivariate normal probability distribution to fit certain linear constraints, described in Subsection~\ref{subsec:effective FI} of the main text, is illustrated by Fig.~\ref{fig:iterations}. The modification of the probability distribution follows the ideas from Ref.~\cite{mikhalychev2015bayesian}, but with certain changes discussed below.

After the coordinate transformation at the $i$-th iteration, one ends up with the probability distribution
\begin{equation}
    \label{eq:app:distribution}
    p^{(i)}(\boldsymbol{\theta}') \propto \exp \left[ -\frac{1}{2}(\boldsymbol{\theta}')^T \boldsymbol{\theta}' \right].
\end{equation}
For the $k$-th constraint $(\mathbf{a}_k')^T \boldsymbol{\theta}' \le b_k'$, one can define the probability of its violation
\begin{equation}
    \label{eq:app:violation probability}
    P_k = \int_{(\mathbf{a}_k')^T \boldsymbol{\theta}' > b_k'} p^{(i)}(\boldsymbol{\theta}') d\boldsymbol{\theta}' = \frac{1}{2}\left[ 1 - \operatorname{erf}\left( \frac{b_k'}{\sqrt 2 |\mathbf{a}_k'|} \right) \right].
\end{equation}
The index of the most severely violated constraint is defined as
\begin{equation}
    j = \argmax_k P_k = \argmin_k \frac{b_k'}{|\mathbf{a}_k'|}.
\end{equation}

If the maximal probability of constraint violation $P_j$ is smaller than certain threshold value (0.01 for the presented numerical calculations), the iterations are stopped and the probability distribution $p^{(i)}(\boldsymbol{\theta})$ is treated as the final result. Otherwise, one can introduce the variable $x = \mathbf{d}^T \boldsymbol{\theta}'$ (where $\mathbf{d} = \mathbf{a}_j' / |\mathbf{a}_j'|$), use the symmetry of the distribution (\ref{eq:app:distribution}), and consider the marginal distribution $w'(x) \propto \exp( - x^2 / 2)$ with the constraint $x \le x_0 \equiv b_j' / |\mathbf{a}_j'| $. Then, the considered iteration modifies the marginal distribution as
\begin{equation}
    w'(x) \mapsto w''(x) \propto \exp\left[-\frac{1 + \xi}{2} (x + \delta)^2 \right]
\end{equation}
with the parameters $\xi$ and $\delta$ being chosen according to the following requirements:
\begin{itemize}
    \item The probability of the constraint violation should be reduced by certain quantity $\eta$ (small enough to ensure convergence of iterations --- $\eta = 0.1$ for the presented numerical calculations): $P_j \mapsto P_j' = \max(P_j / 2, P_j - \eta)$:
    \begin{equation}
        P_j' = \frac{1}{2}\left[ 1 - \operatorname{erf}\left( \frac{(x_0 + \delta) \sqrt{1 + \xi}}{\sqrt 2} \right) \right].
    \end{equation}
    \item The width of the distribution in the prescribed domain $x \le x_0$ should be preserved: \begin{equation}
        \operatorname{Var}(x | w'(x), x \le x_0) = \operatorname{Var}(x | w''(x), x \le x_0).
    \end{equation}
\end{itemize}
The latter requirement differs from the one considered in Ref.~\cite{mikhalychev2015bayesian}, where preservation of the mean value $\operatorname{E}(x | w'(x), x \le x_0) = \operatorname{E}(x | w''(x), x \le x_0)$ was more appropriate for the quantum state reconstruction task.

The two listed requirements imply the following expressions for the quantities $\xi$ and $\delta$:
\begin{equation}
    \label{eq:app:correction xi}
    \xi = \frac{V(P_j', x_0')}{V(P_j, x_0)} - 1,
\end{equation}
\begin{equation}
    \delta = \frac{x_0'}{\sqrt{1 + \xi}} - x_0,
\end{equation}
where $x_0' = \sqrt{2} \operatorname{erf}^{-1}(1 - 2 P_j')$ and
\begin{equation}
    V(p,x) = 1 - \frac{xe^{-x^2/2}}{\sqrt {2 \pi} (1-p)}  - \frac{e^{-x^2}}{2 \pi (1-p)^2}.
\end{equation}

\section{Examples of width estimation by the second derivative with regularization}
\label{app:width estimation}

For the model profile $y_1(x)$, defined in the main text, the second derivative of the logarithm takes the values
\begin{equation}
    Y^{(2)}(x') = \left[ \begin{array}{cc}
        0 & \text{ for } |x'| < x_0 \\
        1/\sigma^2 & \text{ for } |x'| \ge x_0
    \end{array}\right.
\end{equation}
Therefore, according to the definition (\ref{eq:shifted width}), $\Delta(x') \rightarrow \infty$ for $x' < x_0$ and $\Delta(x') = |x'| + \sigma$ otherwise. The minimum of $\Delta(x')$ is reached at $x' = \pm x_0$ and yields the accurate estimate of the half-width
\begin{equation}
    \sigma_\text{min} = x_0 + \sigma.
\end{equation}

For the profile $y_2(x)$, the quantity $Y^{(2)}(x')$ equals
\begin{equation}
    Y^{(2)}(x') = - \frac{k(k-1) {|x'|}^{k-2}}{2 \sigma^k}.
\end{equation}
The displaced half-width $\Delta(x')$, defined by Eq.~(\ref{eq:shifted width}) is equal to 
\begin{equation}
    \Delta(x') = |x'| + \sqrt{\frac{2 \sigma^k}{k (k-1)}} {|x'|}^{1 - k/2}
\end{equation}
and reaches the minimal value at
\begin{equation}
    x' = \pm \left[ \frac{(k-2)^2}{2 k (k - 1)} \right]^{1/k} \sigma.
\end{equation}
The resulting half-width estimate is
\begin{equation}
    \sigma_\text{min} = \left[ \frac{(k-2)^2}{2 k (k - 1)} \right]^{1/k} \frac{k}{k-2} \sigma.
\end{equation}
The dependence of the ratio $\sigma_\text{min} / \sigma$ on $k$ is shown in Fig.~\ref{fig:width regularization}.

\section{Fisher information for 1-parameter problem with discrete and continuous signal}
\label{app:discretness}
Here, we show that discretness 
of the detected signal strongly influences the behavior of FI for dark objects. For simplicity, the 1-parameter model for a uniform object with the transmission amplitude $A$ is considered and the mean value of the detected signal is assumed to scale as $S(A) \propto A^{2n}$.

First, let us consider the typical case, when the detected signal $Y$ takes integer values distributed according to the Poisson distribution (\ref{eq:Poisson likelihood}) with the mean $S(A)$. According to Eq.~(\ref{eq:FIM Poisson}), the FI equals
\begin{equation}
    F_\text{P} = \frac{1}{S(A)}\left( \frac{dS(A)}{dA} \right)^2 \propto A^{2(n-1)}.
\end{equation}

Now, let us approximate the Poisson distribution by a continuous-variable Gaussian distribution with the same mean and variance:
\begin{equation}
    L_\text{G}(Y|S(A)) = \frac{1}{\sqrt{2 \pi S(A)}}\exp\left(- \frac{(Y - S(A))^2}{2 S(A)} \right).
\end{equation}
Then, FI can be calculated as
\begin{multline}
    F_\text{G} = \int_{-\infty}^\infty dY \frac{1}{L(Y|S(A))} \left( \frac{dL(Y | S(A))}{dA} \right)^2 \\ = \left(\frac{1}{S(A)} + \frac{1}{2 S^2(A)} \right) \left( \frac{dS(A)}{dA} \right)^2
\end{multline}

In the limit of strong signal (bright object) $S(A) \gg 1$, the two noise distributions yield the same FI: $F_\text{G} \approx F_\text{P}$. However, for weak signal (dark object), the asymptotics is different: 
\begin{equation}
    F_\text{G} \sim \frac{1}{2 S^2(A)} \left( \frac{dS(A)}{dA} \right)^2 \propto A^{-2} \rightarrow \infty
\end{equation}
while $F_\text{P} \rightarrow 0 $ for $A \rightarrow 0$ and $n > 1$.

\section{Influence of problem parametrization on Fisher information}
\label{app:parameterization}

\subsection{Parametrization-dependence of Fisher information}

The toy problem, introduced in Subsection~\ref{subsec:1D}, can also be parameterized by the transmission probability $T = A^2$. In that case, the FI, calculated according to Eq.~(\ref{eq:FIM Poisson}) for such parametrization, will be equal to 
\begin{equation}
    \label{eq:app:FIM 1D T}
    \tilde F(T) = n^2 N \eta^n T^{n-2} \propto A^{2(n-2)}.
\end{equation}
Comparing Eqs.~(\ref{eq:FIM 1D}) and (\ref{eq:app:FIM 1D T}), one can notice that the FI has different asymptotics for $A\rightarrow 0$: if $n = 2$, FI for the transmission amplitude tends to zero, $F(A) \propto A^{2(n-1)} \rightarrow 0$, while  $\tilde F(T) = n^2 N \eta^n$ remains constant. From that perspective, the following questions arise:
\begin{enumerate}
    \item To what extent is FI parametrization-dependent?
    \item Does re-parametrization of a problem help to resolve the issue of the predicted error divergence in practice?
\end{enumerate}

For a 1-parameter problem, one can define the following change of variables:
\begin{equation}
    \label{eq:app:new 1-param}
    \tilde \theta = g(\theta).
\end{equation}
According to Eq.~(\ref{eq:FIM_score}) and the chain rule for derivatives, the FI for the new parameter $\tilde \theta$ equals
\begin{equation}
    \label{eq:app:FIM re-parameterized 1D}
    \tilde F  = \frac{1}{J^2} F,\quad J = \frac{d \tilde \theta}{d \theta} = \frac{dg(\theta)}{d\theta}.
\end{equation}
CRB for the new parameter $\tilde \theta$ takes the form
\begin{equation}
    \label{eq:app:error re-parameterized 1D}
    \Delta \tilde \theta ^2 \ge 1 / \tilde F = J^2 / F.
\end{equation}

If FI tends to zero as $F = \Theta(\theta^{2m})$ for $\theta \rightarrow 0$ (in D.~Knuth's notations \cite{knuth1976big}), one can choose the new parameter by setting $g(\theta) = \theta^{m+1}$ and ensure that the FI $\tilde F$ would scale as $\tilde F = \Theta(1)$ and the error $\Delta \tilde \theta$ would remain finite for $\theta \rightarrow 0$. For FI, defined by Eq.~(\ref{eq:FIM 1D}), the parametrization $\tilde \theta = A^n$ leads to $\tilde F = 4 N \eta^n = \text{const}$ and resolves the FI singularity at $A = 0$. 

Another important example is the so-called ``Rayleigh's curse'' --- scaling of FI as $F = \Theta(d^2)$ for estimation of a small separation $d$ of a pair of point-like incoherent light sources by direct imaging \cite{bettens1999model,tsang2016quantum,paur2018tempering,len2020resolution}. For the parametrization $\tilde \theta = d^2$, one has $\tilde F = \Theta(1)$ and the error divergence does not appear (the inaccuracy $\Delta(d^2)$ remains finite for $d\rightarrow 0$).

The presented considerations can also be extended to multi-parameter problems \cite{albarelli2020perspective,demkowicz2020multi}. The change of variables can be introduced as
\begin{equation}
    \tilde{\boldsymbol{\theta}} = \mathbf{g}(\boldsymbol{\theta}).
\end{equation}
The FIM for the new parametrization can be calculated according to Eq.~(\ref{eq:FIM_score}) and the chain rule for derivatives:
\begin{equation}
    \tilde F = (J^{-1})^T F J^{-1},
\end{equation}
where $J$ is the Jacobian matrix, $J_{\mu\nu} = \partial \tilde \theta_\mu / \partial \theta_\nu$. CRB for the parameters $\tilde{\boldsymbol{\theta}}$ takes the form
\begin{equation}
    \label{eq:app:error re-parameterized multi}
    \operatorname{Cov}(\tilde{\boldsymbol{\theta}},\tilde{\boldsymbol{\theta}}) \ge \tilde F^{-1} = J F^{-1} J^T.
\end{equation}

To conclude, the problem parametrization influences the behavior of FIM significantly and can change its asymptotics. Therefore, a methodological ``chicken-or-egg'' question arises: What comes first --- the measurement or the problem parametrization? Do we seek for \textit{any} kind of information about the investigated object \textit{after} the measurement has been performed? Or do we specify the measurement \textit{purpose} (the set of parameters $\boldsymbol{\theta}$ to be inferred) at the stage of just \textit{planning} the experiment? For practical applications, the latter approach is generally favored. Some remarks on the optimal choice of parameters are presented in Appendix~\ref{app:subsec:parameterization}.

\subsection{Fisher information and error analysis for a predefined set of parameters.}

In the main text, we do not consider the task of finding the optimal task-specific parametrization and assume that the problem statement already prescribes certain fixed ``natural'' set of parameters $\boldsymbol{\theta}$ and enforces us to report the results in terms of those specific predefined variables. After estimation of the errors for the introduced parameters $\tilde{\boldsymbol{\theta}}$ according to Eqs.~(\ref{eq:app:error re-parameterized 1D}) and (\ref{eq:app:error re-parameterized multi}), one needs to return to the initial set of parameters $\boldsymbol{\theta}$. According to the error propagation rules, the errors for the initial variables can be estimated as
\begin{equation}
    \label{eq:app:error return 1D}
    \Delta \theta^2 = \Delta \tilde \theta^2 / J^2
\end{equation}
for a 1-parameter model and 
\begin{equation}
    \label{eq:app:error return multi}
    \operatorname{Cov}(\boldsymbol{\theta},\boldsymbol{\theta}) = J^{-1} \operatorname{Cov}(\tilde{\boldsymbol{\theta}},\tilde{\boldsymbol{\theta}}) (J^{-1})^T
\end{equation}
in general case.

The transition from $F = \Theta(\theta^{2m}) \rightarrow 0$ for $\theta \rightarrow 0$ to $\tilde F = \Theta(1)$, discussed above, requires $J = \Theta(\theta^m) \rightarrow 0$. Therefore, the problem of the error divergence re-appears as soon as we return to the initial variable(s). For example, in the case of ``Rayleigh's curse'', a small error $\Delta (d^2)$ of the squared separation $d^2$ does not necessarily imply accurate estimation of the separation $d$ itself for small $d$, since $\Delta d \approx \Delta(d^2) / (2d)$. Similarly, a small error of $\tilde \theta = A^n$ in the toy example from Subsection~\ref{subsec:1D} does not guarantee accurate inference of the transmission amplitude $A$.

As a result, the re-parametrization of the problem just shifts the necessity of regularization from the FIM $F$ to the errors recalculation according to Eqs.~(\ref{eq:app:error return 1D}) and (\ref{eq:app:error return multi}). For a 1-parameter case, one can go beyond the linear approximation lying behind the error propagation rules and estimate $\Delta \theta$ in a more accurate way (for example, by re-calculating Bayesian posterior probabilities). In a general multiparametric case, the problem of regularizing Eq.~(\ref{eq:app:error return multi}) looks at least as complicated as our approach operating with initial FIM $F$ itself without resorting to the re-parametrization.

\subsection{Optimal parametrization for Fisher information analysis}
\label{app:subsec:parameterization}

In this section, we briefly discuss how the choice of the problem parametrization can improve calculation of FI and application of CRB, if no strict constraints are imposed on the object parameters to be reported. 

If the problem is described by a single parameter $\bar \theta$, defined by Eq.~(\ref{eq:app:new 1-param}) and known \textit{a priori} to have the value close to $\bar \theta_0$, one can construct the efficient (saturating CRB) locally unbiased estimator \cite{demkowicz2020multi,helstrom1969quantum}:
\begin{multline}
    \label{eq:app:locally unbiased 1-param}
    \hat{\bar \theta}(Y, \bar \theta_0) = \bar \theta_0 +\bar F^{-1} \left.\frac{d}{d\bar \theta} \log L(Y|S(\bar \theta))\right|_{\bar \theta = \bar \theta_0} \\ = \left( \bar \theta_0 -  \frac{S(\bar \theta_0)}{S'(\bar \theta_0)} \right) + \frac{Y}{S'(\bar \theta_0)},
\end{multline}
where we took into account Eq.~(\ref{eq:Poisson likelihood}) and introduced the notation $S'(\bar \theta_0) = \left. dS(\bar \theta) / d \bar \theta \right|_{\bar \theta = \bar \theta_0} $. Here, we do not take into account the additional bias imposed by the upper bound ($A \le 1$ in the toy example in Subsection~\ref{subsec:1D}).

If, for certain parametrization (specific choice of $\bar \theta$), both terms in the right-hand side of Eq.~(\ref{eq:app:locally unbiased 1-param}) do not depend on $\bar \theta_0$, $\hat{\bar \theta}(Y)$ represents the globally unbiased efficient estimator:
\begin{equation}
    \label{eq:app:eqs for globally unbiased}
    \begin{gathered}
        S'(\bar \theta_0) = \alpha,\\
        \frac{S(\bar \theta_0)}{S'(\bar \theta_0)} - \bar \theta_0 = \beta,
    \end{gathered}
\end{equation}
where $\alpha$ and $\beta$ are constants (independent of $\bar \theta_0)$. The system of equations (\ref{eq:app:eqs for globally unbiased}) implies that the relation between the signal $S$ and the parameter $\bar \theta$ must be linear:
\begin{equation}
    S(\bar \theta) =  \alpha \bar \theta + \beta.
\end{equation}
In particular, for the 1-parameter toy example, $\bar \theta$ can be chosen as $\bar \theta = A^{2n}$. It is worth noting that the constructed unbiased estimator $\hat{\bar \theta}(Y)$ coincides with MLE for that parametrization.

In contrast to the optimal parametrization, the initial parametrization of the toy problem by the transmission amplitude $A$ does not allow construction of a globally unbiased estimator. Since $S(0) = 0$, the mean bias of an estimator $\bar A(Y)$ at the point $A = 0$ equals
\begin{equation}
   \left. \langle \bar A \rangle \right|_{A = 0} = \sum_{Y = 0}^\infty \bar A(Y) L(Y|S(0)) = \bar A(0).
\end{equation}
Therefore, only estimators with $\bar A(0) = 0$ are unbiased at $A = 0$. The derivative of the mean bias at $A = 0$ for the toy example equals
\begin{multline}
    \left. \frac{d \langle \bar A \rangle}{dA} \right|_{A = 0} = \bar A(1) \left. \frac{d S(A)}{dA} \right|_{A = 0} \\ = \left. 2n \bar A(1) N \eta^n A^{2n - 1} \right|_{A = 0} = 0
\end{multline}
for $n \ge 1$, while for an unbiased estimator the equality
\begin{equation}
    \frac{d \langle \bar A \rangle}{dA} = 1
\end{equation}
must hold.

Treatment of a general multiparametric case is similar. An efficient locally unbiased estimator is constructed as \cite{demkowicz2020multi}
\begin{multline}
    \label{eq:app:locally unbiased multi-param}
    \hat{\bar {\boldsymbol{\theta}}}(\mathbf{Y}, \bar { \boldsymbol{\theta}}_0) = \bar { \boldsymbol{\theta}}_0 +\bar F^{-1} \left.\boldsymbol{\nabla} \log L(\mathbf{Y}|\mathbf{S}(\bar { \boldsymbol{\theta}}))\right|_{\bar { \boldsymbol{\theta}} = \bar { \boldsymbol{\theta}}_0} \\ = \bar { \boldsymbol{\theta}}_0 + \sum_i F^{-1}(\bar { \boldsymbol{\theta}}_0)\left( \frac{Y_i}{S_i(\bar { \boldsymbol{\theta}}_0)} - 1 \right) \left.\boldsymbol{\nabla} S_i(\bar { \boldsymbol{\theta}})\right|_{\bar { \boldsymbol{\theta}} = \bar { \boldsymbol{\theta}}_0} \\ \equiv C(\bar { \boldsymbol{\theta}}_0) \mathbf{Y} + \mathbf{D}(\bar { \boldsymbol{\theta}}_0),
\end{multline}
where 
\begin{equation}
    \label{eq:app:estimator D expression}
    \mathbf{D}(\bar { \boldsymbol{\theta}}_0) = \bar { \boldsymbol{\theta}}_0 - C(\bar { \boldsymbol{\theta}}_0) \mathbf{S}(\bar { \boldsymbol{\theta}}_0).
\end{equation}
The estimator is globally unbiased and efficient if the matrix $C(\bar { \boldsymbol{\theta}}_0) = C$ and the vector $\mathbf{D}(\bar { \boldsymbol{\theta}}_0) = \mathbf{D}$ do not depend on the parameters' value $\bar { \boldsymbol{\theta}}_0$. In that case Eq.~(\ref{eq:app:estimator D expression}) implies that such estimator can be constructed only for a parametrization with the following linear relation between the parameters vector $\bar { \boldsymbol{\theta}}$ and the signal $\mathbf{S}(\bar { \boldsymbol{\theta}})$:
\begin{equation}
    \label{eq:app:parametrization linearity}
    \bar { \boldsymbol{\theta}} = C \mathbf{S}(\bar { \boldsymbol{\theta}}) + \mathbf{D}.
\end{equation}

For the 2-parameter toy example, discussed in Subsection~\ref{subsec:2D}, the parameters choice
\begin{equation}
    \bar { \theta}_1 = S_1(\mathbf{A}),\quad \bar \theta_2 = S_2(\mathbf{A})
\end{equation}
indeed ensures that the MLE is globally unbiased and efficient (except for vicinity of the upper bound $A_i \le 1$). However, such parametrization is far from being practical. The ultimate goal of imaging (microscopy) is to retrieve \textit{local} information about the object features from the results of available measurements. Even allowing nonlinear re-parametrization, one would prefer to retain locality by defining the $i$-th new parameter as a function of the $i$-th initial parameter only:
\begin{equation}
    \label{eq:app:re-parametrization locality}
    \bar \theta _i = g_i (\theta_i).
\end{equation}
For the considered 2-parameter toy example, Eqs.~(\ref{eq:app:parametrization linearity}) and (\ref{eq:app:re-parametrization locality}) can be satisfied simultaneously if the equality
\begin{equation}
    2 h_0 h_1 (h_1^2 - h_0^2) = 0
\end{equation}
holds, which is possible only in trivial special cases: $h_0 = 0$, or $h_1 = 0$ (no blurring), or $h_0 = h_1$ (degenerate problem with $S_1(\mathbf{A}) \equiv S_2(\mathbf{A})$).

For the multiparametric problems, discussed in Subsection~\ref{subsec:practical_applications}, an attempt to satisfy the conditions of linearity (Eq.~(\ref{eq:app:estimator D expression})) and locality (Eq.~(\ref{eq:app:re-parametrization locality})) simultaneously leads to the same issue: the number of additional constraints imposed on the imaging system parameters $D^{(ij)}_{ml}$ grows quadratically relatively to the number of the object parameters $A_i$ (and $\bar \theta_i$). Therefore, the existence of the optimal parametrization allowing construction of the globally unbiased efficient estimator and satisfying the locality restriction (\ref{eq:app:re-parametrization locality}) is limited to certain very special cases. 

The necessity to deal with parametrizations not allowing construction of globally unbiased efficient estimators is common for quantum multi-parameter estimation problems \cite{demkowicz2020multi,albarelli2020perspective,yang2019attaining,barndorff2000fisher}. In the limit of a large number of detected events (measurement repetitions), the MLE becomes asymptotically unbiased (except for the parameters belonging to the border of the physical region $\Omega$) and CRB provides an accurate estimate of error \cite{demkowicz2020multi,yang2019attaining,albarelli2020perspective}. The effect manifests itself as a small bias in the region II in Fig.~\ref{fig:1D results} even for the relatively small mean number of detection events $N \eta^n = 98$.

\begin{figure}
    \centering
    \includegraphics[width=0.8\linewidth]{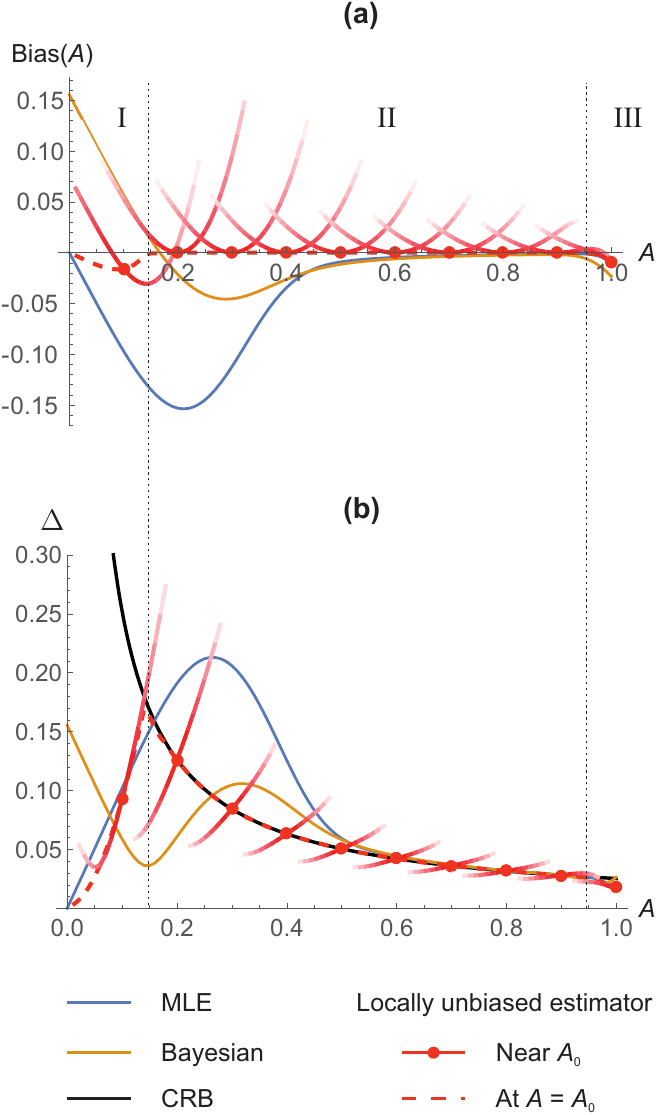}
    \caption{Transmission amplitude estimation for a uniform object (single-parameter model): estimation bias (a) and MSE (b). Solid red lines show the behavior of the locally unbiased estimator $\hat A(Y, A_0)$, defined by Eq.~(\ref{eq:app:locally unbiased 1-param explicit}), for 10 values $A_0$ indicated by red dots. Dashed red lines aggregate the bias and MSE for the estimators $\hat A(Y, A_0)$ with the value $A_0$ equal to the true value $A$. Other lines are taken from Fig.~\ref{fig:1D results}. The regions I--III are defined according to the bias of the estimators $\hat A(Y, A_0)$ with $A_0 = A$ (biased in I and III, locally unbiased in II).}
    \label{fig:app:1D results}
\end{figure}

If the number of detection events is not large, one can use the locally unbiased estimator \cite{helstrom1969quantum,demkowicz2020multi}, defined by Eqs.~(\ref{eq:app:locally unbiased 1-param}) and (\ref{eq:app:locally unbiased multi-param}), with the rough estimate of $\bar { \boldsymbol{\theta}}_0$ being found by the procedure outlined in Ref.~\cite{barndorff2000fisher}. For the 1-parameter toy example with the initial parametrization by the transmission amplitude $A$, the resulting estimator takes the form
\begin{equation}
    \label{eq:app:locally unbiased 1-param explicit}
    \hat A (Y, A_0) = \min \left\{1, A_0 \left[1 + \frac{1}{2 n} \left(\frac{Y}{S(A_0)} - 1\right) \right] \right\},
\end{equation}
where the upper bound $A \le 1$ is taken into account explicitly. The bias and MSE of that estimator for the parameters, used in Fig.~\ref{fig:1D results}, are shown in Fig.~\ref{fig:app:1D results}. As expected, the constructed estimator is locally unbiased and saturates CRB in the region II, not including the vicinity of the boundary values $A=0$ and $A=1$. The constraint, introduced by the function $\min\{1,\ldots\}$ in Eq.~(\ref{eq:app:locally unbiased 1-param explicit}), leads to the estimation bias and decrease of MSE relatively to CRB in the regions I and III. To account for such effect, the FI modification and regularization approach, proposed in the current paper, can be applied. It is interesting to note that the bias in region I is still caused by the \textit{upper} bound $A \le 1$: the locally unbiased estimator $\hat A (Y, A_0)$, defined by Eq.~(\ref{eq:app:locally unbiased 1-param}), diverges for $A_0 \rightarrow 0$ ($\lim_{A_0 \rightarrow 0} \hat A(1, A_0) = \infty$), while the constraint introduced in Eq.~(\ref{eq:app:locally unbiased 1-param explicit}) regularizes it at the cost of additional bias.

\section{Calculations for the 1-parameter problem of transmission amplitude estimation}
\subsection{Optimal biased estimate}
\label{app:optimal bias}

Optimizing the bias of a parameter estimate, one can reduce the MSE of the parameter relatively to unbiased estimates \cite{benHaim2009lower,eldar2004minimum,eldar2006uniformly,eldar2008rethinking}. For the model, described in Subsection~\ref{subsec:1D}, one can specify the cost function for the bias optimization as the MSE averaged over the whole physical range of the true parameter values $A \in [0,1]$:
\begin{equation}
    \label{eq:app:mean MSE}
    \langle \Delta^2 \rangle_A \equiv \int_0^1 dA \operatorname{E}[(\hat A(Y) - A)^2]
\end{equation}
where $\hat A(Y)$ is the estimator to be optimized, and the expectation value over the signal realization is defined as
\begin{equation}
    \label{eq:app:expectation definition}
    \operatorname{E} [f(Y)] = \sum_{Y = 0}^\infty L(Y | S(A)) f(Y).
\end{equation}

Substituting Eq.~(\ref{eq:app:expectation definition}) into Eq.~(\ref{eq:app:mean MSE}), one can obtain the following expression for the cost function:
\begin{multline}
    \langle \Delta^2 \rangle_A = \sum_{Y = 0}^\infty \Bigl[ \langle L(Y|S(A)) \rangle_A \hat A^2(Y) \\ - 2\langle L(Y|S(A)) A \rangle_A \hat A(Y) + \langle L(Y|S(A) A^2) \rangle_A \Bigr],
\end{multline}
where the brackets $\langle \cdots \rangle$ denote averaging over $A$ as in Eq.~(\ref{eq:app:mean MSE}). The optimization is performed over the discrete set of variables $\hat A(Y)$, parameterized by an integer index $Y$:
\begin{equation}
    \frac{\partial}{\partial \hat A(Y)} \langle \Delta^2 \rangle_A = 0 \text{ for all }Y = 0, 1, \ldots.
\end{equation}
The minimum is reached when the following equation is satisfied:
\begin{equation}
    \hat A(Y) = \frac{\langle L(Y|S(A)) A \rangle_A}{\langle L(Y|S(A)) \rangle_A} = \frac{\int_0^1 dA' L(Y|S(A')) A'}{\int_0^1 dA' L(Y|S(A'))}.
\end{equation}
The result coincides with the mean \textit{a posteriori} estimate defined by Eq.~(\ref{eq:Bayesian 1D definition}).

\subsection{Mean squared error estimation}
\label{app:MSE for 1D}

For numerical analysis of the estimation error in a 1-parameter case, one can start from constructing the mapping dictionary $Y \mapsto \hat A(Y)$ for the considered estimate. Eqs.~(\ref{eq:MLE 1D definition}) and (\ref{eq:Bayesian 1D definition}) are to be used for MLE and Bayesian estimates defined in the main text. Then, the MSE and the bias are calculated as functions of the true parameter value $A$ as
\begin{equation}
    \operatorname{MSE}(A) = \operatorname{E}[(\hat A(Y) - A)^2], \; \operatorname{Bias}(A) = \operatorname{E}[\hat A(Y) - A],
\end{equation}
where the expectation value is defined by Eq.~(\ref{eq:app:expectation definition}).

FI-based error evaluation includes application of standard CRB ($\operatorname{Var}(A) \sim 1 / F$), modified CRB for a biased estimator \cite{eldar2004minimum,eldar2008rethinking}
\begin{multline}
    \operatorname{MSE}(A) = \operatorname{Var}(A) + [\operatorname{Bias}(A)]^2 \\ \sim \left( 1+ \frac{\partial \operatorname{Bias}(A)}{\partial A}\right)^2 \frac{1}{F} + [\operatorname{Bias}(A)]^2,
\end{multline}
and the proposed approach with the corrected FI and standard CRB ($\operatorname{MSE} \sim 1 / \tilde F$).

\subsection{Fisher information regularization for dark objects}
\label{app:regularization 1D}

Regularization of FI for small $A$ is performed according to Eq.~(\ref{eq:regularized FIM 1D}) after substitution of Eq.~(\ref{eq:FIM 1D}):
\begin{equation}
    \label{eq:app:regularized FIM 1D}
    \tilde F(A) = \max_{A'} \frac{K {A'}^{2(n-1)}}{\left(1 + |A' - A| \sqrt{K} {A'}^{n-1}\right)^2}
\end{equation}
where $K = 4 n^2 N \eta^n$. For $A' < A$ the derivative of the optimized expression is positive, i.e. the optimal $A'$ is not less than $A$. For $A' > A$ the maximum of the optimized expression is reached at 
\begin{equation}
    A'_\text{opt} = \left( \frac{n-1}{\sqrt K} \right)^{1/n}
\end{equation}
if $A \le A'_\text{opt}$. Otherwise, the optimal value is $A' = A$ and no regularizing correction is applied.

\subsection{Fisher information correction for the constraints}
\label{app:correction 1D}

For the 1-parameter model, considered in Subsection~\ref{subsec:1D}, the constraints to be taken into account by FI modification are $0 \le A \le 1$. The input FI is the result of regularization described in Appendix~\ref{app:regularization 1D}. To avoid confusion with the final result of correction, we denote the input FI as $F$ (not $\tilde F$ as in Eq.~(\ref{eq:app:regularized FIM 1D})). We follow the procedure from Subsection~\ref{subsec:effective FI}.

Both, the FIM $F$ and the transformation matrix $T$ have dimensions $1 \times 1$ in the considered case and represent just scalars: $T = \sqrt F$. The parameter transformation (for the first iteration) is $A' = \sqrt F (A - A^{(0)})$, where $A^{(0)}$ is the transmission amplitude value, for which FI is analyzed. Parameterization of the constraints in the form $a_j A \le b_j$ is
\begin{equation}
    A \le 1 \;\Rightarrow\; a_1 = 1,\; b_1 = 1,
\end{equation}
\begin{equation}
    A \ge 0 \;\Rightarrow\; -A \le 0 \;\Rightarrow\; a_0 = -1,\; b_0 = 0.
\end{equation}
The transformed parameters of the constraints are
\begin{equation}
    a_1' = \frac{1}{\sqrt F},\; b_1' = 1 - A^{(0)},\; a_0' = - \frac{1}{\sqrt F},\; b_0' = A^{(0)}.
\end{equation}
The local variable $x$, introduced in Appendix~\ref{app:Iterative algorithm}, coincides with the transformed parameter up to the sign: $x = \pm A'$, where the ``+'' and ``$-$'' signs correspond to the constraint indices 1 and 0 respectively. The limiting values $x_0^{(j)} = b_j / |a_j|$ for the variable $x$ are equal to
\begin{equation}
    x_0^{(1)} = \sqrt{F} \left(1 - A^{(0)}\right),\; x_0^{(0)} = \sqrt{F} A^{(0)}.
\end{equation}

The probabilities $P_1$ and $P_0$ of the constraints violation are defined by Eq.~(\ref{eq:app:violation probability}). For the model, discussed in Subsection~\ref{subsec:1D}, the following relations hold for those probabilities:
\begin{equation}
    P_1 \le P_\text{threshold} \equiv 0.01\text{ for } A \le 0.937
\end{equation}
and
\begin{equation}
    P_0 \le P_\text{threshold} \text{ for } A \ge 0.242.
\end{equation}
Therefore, for each $A$ at most one of the constraints is active and requires application of the correction procedure. Due to the additivity of corrections applied to the same direction \cite{mikhalychev2015bayesian}, one can avoid multiple iterations and reduce the constraint violation probability from its initial value to the target value $P_\text{target}$ in one step.

The shrinking parameter $\xi$ for the discussed correction step is prescribed by Eq.~(\ref{eq:app:correction xi}) with $P_j' = P_\text{target}$. According to the last step of the correction procedure described in Subsection~\ref{subsec:effective FI}, the resulting corrected FI equals $\tilde F = (1 + \xi) F$, where $F$ is the input FI (namely, the regularization result).

\section{Visualization of Fisher information and covariance matrix for a 2-parameter model}
\label{app:visualization 2D}

Since FIM is a positive semi-definite matrix, it can serve as a quadratic form defining an ellipse in a 2-parameter case $\boldsymbol{\theta}= (\theta_1, \theta_2)$. The meaning of the ellipse is clearly seen from Eqs.~(\ref{eq:quadratic form covariance}) and (\ref{eq:quadratic form covariance approx}) (see also Fig.~1 in Ref.~\cite{mikhalychev2021fisher} for illustration). If the estimation results are scattered according to multivariate normal distribution
\begin{equation}
    \label{eq:app:quadratic form FIM}
    p(\boldsymbol{\theta}) = \frac{ \sqrt{\operatorname{det} F}}{2 \pi} \exp\left[ - \frac{1}{2} \Delta \boldsymbol{\theta}^T F \Delta \boldsymbol{\theta} \right],\quad \Delta \boldsymbol{\theta} = \boldsymbol{\theta} - \boldsymbol{\theta}_0, 
\end{equation}
centered at some point $\boldsymbol{\theta}_0$, the equation
\begin{equation}
    \Delta \boldsymbol{\theta}^T F \Delta \boldsymbol{\theta} = 2 \log 2
\end{equation}
defines the boundary between the regions with $p(\boldsymbol{\theta}) > 1/2$ and $p(\boldsymbol{\theta}) < 1/2$. Moreover, each region will, in average, contain a half of the sampled results:
\begin{equation}
    \int\limits_{\Delta \boldsymbol{\theta}^T F \Delta \boldsymbol{\theta} \le 2 \log 2} d^2 \boldsymbol{\theta} p(\boldsymbol{\theta}) = \frac{1}{2 \pi} \int\limits_{|\mathbf{r}|\le 1} d^2 \mathbf{r} e^{-\mathbf{r}^2 / 2} = \frac{1}{2}.
\end{equation}

Since CRB with standard FIM is valid for unbiased estimates, it is reasonable to take $\boldsymbol{\theta}_0$ equal to the true value of the parameters $\boldsymbol{\theta}$ for visualization of the standard FIM. The correction procedure (Subsection~\ref{subsec:effective FI}) shifts the probability distribution. Therefore, $\boldsymbol{\theta}_0$ should be taken as the center $\boldsymbol{\theta}^{(i_\text{max})}$ of the distribution after the last iteration (with the index $i_\text{max}$) for visualization of the corrected FIM $\tilde F$.

To visualize statistics of a set of randomly sampled points (in 2-dimensional space), one can assume a multivariate normal distribution based on the sample mean $\langle \boldsymbol{\theta} \rangle$ and covariance $C = \operatorname{Cov}(\boldsymbol{\theta}, \boldsymbol{\theta})$:
\begin{equation}
    p(\boldsymbol{\theta}) = \frac{1}{2 \pi  \sqrt{\operatorname{det} C}} \exp\left[ - \frac{1}{2} \Delta \boldsymbol{\theta}^T C^{-1} \Delta \boldsymbol{\theta} \right],\; \Delta \boldsymbol{\theta} = \boldsymbol{\theta} - \langle \boldsymbol{\theta} \rangle.
\end{equation}
The quadratic form of the distribution corresponds to the ellipse defined by the equation
\begin{equation}
    \Delta \boldsymbol{\theta}^T C^{-1} \Delta \boldsymbol{\theta} = 2 \log 2
\end{equation}
in the sense discussed above.

\end{document}